\documentclass[conf]{new-aiaa}
\usepackage[utf8]{inputenc}

\usepackage{booktabs}

\usepackage{hyperref}

\usepackage{graphicx}
\usepackage{amsmath}
\usepackage[version=4]{mhchem}
\usepackage{siunitx}
\usepackage{longtable,tabularx}
\usepackage[ruled,vlined]{algorithm2e}

\DeclareMathOperator*{\minimize}{minimize}

\title{Attitude Trajectory Optimization for Agile Satellites in Autonomous Remote Sensing Constellations}

\author{Emmanuel Sin\footnote{Graduate Student, Department of Mechanical Engineering, emansin@berkeley.edu, AIAA Student Member.}, Murat Arcak\footnote{Professor, Department of Electrical Engineering and Computer Sciences, arcak@berkeley.edu.}}
\affil{University of California, Berkeley, California, 94720, U.S.A.}
\author{Sreeja Nag\footnote{Senior Research Scientist, Bay Area Environmental Research Institute, sreeja.nag@nasa.gov, AIAA Senior Member.}, Vinay Ravindra\footnote{Research Scientist, Bay Area Environmental Research Institute, vinay.ravindra@nasa.gov, AIAA Member.}, Alan Li\footnote{Research Scientist, Bay Area Environmental Research Institute, alan.s.li@nasa.gov.} }
\affil{NASA Ames Research Center, Moffett Field, California, 94035, U.S.A.}

\begin{document}

\maketitle

\begin{abstract}
Agile attitude maneuvering maximizes the utility of remote sensing satellite constellations. By taking into account a satellite's physical properties and its actuator specifications, we may leverage the full performance potential of the attitude control system to conduct agile remote sensing beyond conventional slew-and-stabilize maneuvers. Employing a constellation of agile satellites, coordinated by an autonomous and responsive scheduler, can significantly increase overall response rate, revisit time and global coverage for the mission. In this paper, we use recent advances in sequential convex programming (SCP) based trajectory optimization to enable rapid-target acquisition, pointing and tracking capabilities for a scheduler-based constellation. We present two problem formulations. The \textit{Minimum-Time Slew Optimal Control Problem} determines the minimum time, required energy, and optimal trajectory to slew between any two orientations given nonlinear quaternion kinematics, gyrostat and actuator dynamics, and state/input constraints. By gridding the space of 3D rotations and efficiently solving this problem on the grid, we produce lookup tables or parametric fits off-line that can then be used on-line by a scheduler to compute accurate estimates of minimum-time and maneuver energy for real-time constellation scheduling. The estimates allow an optimization-based scheduler to produce target-remote-sensing and data-downlinking schedules that are dynamically feasible for each satellite and optimal for the constellation. The \textit{Minimum-Effort Multi-Target Pointing Optimal Control Problem} is used on-line by each satellite to produce continuous attitude-state and control-input trajectories that realize a given schedule while minimizing attitude error and control effort. The optimal trajectory may then be achieved by a low-level tracking controller. This onboard trajectory generation and tracking scheme is possible due to real-time, efficient SCP implementations. We demonstrate our approach with a numerical example that uses simulation data for a reference satellite in Sun-synchronous orbit passing over globally-distributed, Earth-observation targets.
\end{abstract}

\section{Nomenclature}

{\renewcommand\arraystretch{1.0}
\noindent\begin{longtable*}{@{}l @{\quad=\quad} l@{}}
$J$                                & mass moment of inertia matrix of rigid body in body-fixed frame, [kg$\cdot$ m$^2$] \\
$J_r$                             & mass moment of inertia of actuation rotor about its axis of rotation, [kg$\cdot$ m$^2$] \\
$A_r$                             & actuator Jacobian, each column representing a rotor's axis of rotation w.r.t. body-fixed frame \\
$\mathbf{q}$                  & unit quaternion parameterizing attitude of rigid body's body-fixed frame w.r.t. inertial frame, [ \ ] \\
$\boldsymbol{\omega}$ & angular velocity of rigid body about body-fixed frame axes,  [rad/s] \\
$\mathbf{r_i}$                & angular momentum of rotor $i$ about its axis,  [Nms] \\
$\mathbf{u}_i$    & torque of rotor $i$ about its rotor axis, [N] \\
$t$				     & time, [s] \\
$\tau$		             & normalized time, [ \ ]
\end{longtable*}}

\section{Introduction}

Due to the proliferation of launch providers to low Earth orbit (LEO) and the trend towards smaller, cost-efficient spacecraft, satellite constellations are enabling scientific missions and commercial applications that are otherwise impossible with a single, larger satellite. For example, a constellation of satellites in LEO may be coordinated to point towards coastal regions around the world to measure ocean color, atmospheric properties, phytoplankton concentrations, and ultimately assess the health of global coral reef ecosystems \cite{Nag1}. LEO constellations may also be employed to measure episodic precipitation and subsequent water flow in flood-prone cities \cite{Nag2}-\cite{Nag3}. Furthermore, constellations may be tasked to measure soil moisture in targeted regions to assess the risk of wildfires \cite{Nag4}. In addition to climate and environment monitoring, Earth-observing constellations are providing commercial and economic value by measuring, for instance, agricultural crop growth, infrastructure development, or logistical activity at airports and shipping container routes \cite{Shah}. Beyond these Earth observation applications, LEO satellite constellations are also enabling space-based Internet and telecommunication services \cite{Harris}.

To enhance performance as an interconnected system, the satellites of a constellation can be precisely coordinated throughout each stage of a mission. Once a group of satellites are deployed into a desired orbital plane by a launch vehicle, the satellites enter the \textit{orbit acquisition stage} where they must be phased relative to each other to achieve desired angular spacing. In \cite{Li}, it is assumed that the orientation of the satellites, with respect to their orbital velocities, may be controlled so that either a minimum or maximum surface area is exposed to the atmosphere that exists in low Earth orbit. By inducing either a low or high atmospheric drag on each satellite using a bang-bang control approach, the resulting differential drag between satellites is used to separate them into an equally-spaced formation. In \cite{Foster}, simulated annealing is used to design time-optimal differential drag commands for a group of up to 100 satellites and the method is demonstrated on an actual constellation deployed in LEO. Assuming accurate attitude pointing, \cite{Sin1} formulates a linear program that produces differential drag commands taking on continuous values between the minimum and maximum values, allowing the constellation to form not only under a minimum-time objective but also with a maximum altitude (or equivalently, maximum constellation lifetime) objective. A distributed controller approach is presented in \cite{Sin2}, where it is assumed that the attitude of the satellites can be controlled to apply continuous low-thrust in the appropriate direction. Regardless of where the control authority is derived from (e.g., differential drag commands or thrust commands), these constellation acquisition methods require accurate attitude pointing. 

In the subsequent \textit{operational stage}, the satellites must perform various scheduled tasks including targeted remote sensing (e.g., imaging, radiometry), downlinking/uplinking data to/from ground stations, orbital station-keeping and other maintenance activities. In \cite{Nag1}-\cite{Nag4}, an automated scheduler is developed to run autonomously on either ground stations (with schedules uplinked to satellites) or onboard in a distributed approach. Based on dynamic programming or mixed integer programming, the scheduler produces imaging schedules for each satellite that maximizes the number of observations and/or observation times for the constellation as a whole. In addition to maximizing spatial imaging coverage, \cite{Shah} also addresses the problem of maximizing data downlinked by the constellation. We note that the schedulers in these works make inherent assumptions on the agility and pointing performance of the attitude control subsystem on board each satellite. For instance, in order to produce a feasible schedule that provides sufficient time to slew between desired orientations, the scheduler must know the dynamically-feasible, minimum slew time between any two desired orientations. This minimum slew time depends on the physical properties of the spacecraft (i.e., mass moment of inertia, actuator configuration) and actuator constraints (e.g., maximum rotor torque and momentum). Once an optimized schedule is produced for the constellation, each satellite must then generate and track an attitude trajectory that realizes its given schedule. Such an attitude trajectory may be optimized to minimize desired-actual attitude error at specific times of the trajectory and minimize control effort over the course of the trajectory. 

Predicting that future spacecraft will require agile attitude control systems that provide rapid multi-target pointing and tracking capabilities, \cite{Wie1} proposes a feedback regulator to conduct large-angle, rest-to-rest slew maneuvers using the Euler's eigenaxis rotation between any two orientations. Resembling feedback linearization, the proposed law introduces a nonlinear term to cancel out the coupling between body angular velocities and replaces it with linear error-quaternion and body rate feedback terms. In \cite{Wen}, a large class of attitude tracking control laws that have the general form of proportional-derivative (PD) feedback and feedforward compensation is obtained with proofs of their global asymptotic stability in the closed-loop. Minimum-time, rest-to-rest slew maneuvers for an inertially symmetric rigid spacecraft with independent three-axis controls are studied in \cite{Bilimoria}, which shows that the optimal maneuver is not, in general, an eigenaxis rotation but one that includes significant nutation of the instantaneous axis of rotation. Furthermore, the structure of the optimal control is different for small and large reorientation angles. While \cite{Wie1}-\cite{Bilimoria} focus on rigid body dynamics under ideal, body-fixed control torques, \cite{Wie2} considers actuator dynamics and presents a feedback control logic that produces near minimum-time eigenaxis slew maneuvers under actuator saturation, slew rate limit, and control bandwidth limit. The control laws are used in \cite{Wie3} to demonstrate rapid multi-ground-target acquisition by stepping through reference set-points that define successive scan trajectories. We note that the attitude control strategy in \cite{Nag1} also uses a minimum-time, eigenaxis slew maneuver and switches to a PD control law for small angles. Based on closed-loop simulations with this control strategy, a polynomial fit of minimum maneuver time as a function of the eigenaxis slew angle is used in a scheduler \cite{Nag2}.

Although eigenaxis-based minimum-time control laws may be applied to general minimum-time problems to produce near-optimal maneuvers, they do not explicitly address nonlinearities that arise from, for instance, actuator dynamics nor do they explicitly consider general state and input constraints (e.g., bounds on momentum, power and energy). Moreover, when directly applying a feedback control strategy to track a desired sequence of discrete orientations, the feedback gains between each pair of orientations must be carefully tuned to achieve settling times that completely satisfy a desired pointing schedule, without missing any targets. In these minimum-time or minimum-attitude-error situations, we may use trajectory optimization methods to not only explicitly deal with nonlinearities and constraints, but to also automate the generation of continuous attitude trajectories for autonomous execution.

As surveyed by Betts \cite{Betts1}, there is an expanse of literature on trajectory optimization methods that may be characterized by, for example, the solution approach (i.e., indirectly satisfying necessary conditions for optimality or directly solving a transcribed version of the problem), or the transcription process (shooting versus collocation). In \cite{Kelly}, various direct collocation methods are introduced while \cite{Ross} focuses on an indirect pseudospectral method that has been used in practice, for example, to control the International Space Station with a zero-propellant maneuver. Regardless of the approach, many trajectory optimization methods treat a problem in its original nonlinear, non-convex form, requiring the use of a nonlinear programming solver \cite{Betts2}, which may be computationally inefficient with long solution times.

Recent advances in sequential convex programming (SCP) have enabled efficient computation of locally optimal trajectories for nonlinear systems with non-convex constraints and objectives. SCP is an iterative method that repeatedly formulates and solves a convex, finite-dimensional parameter optimization problem that approximates the original non-convex optimal control problem. A convex formulation is typically achieved by linearizing the nonlinear system around a nominal trajectory (i.e., the solution from the previous iteration) and approximating any non-convex constraints and objective with Taylor series expansions. Fast and reliable Interior Point Method algorithms \cite{Nocedal} may be used to solve these convex subproblems. In the early works of \cite{Szmuk0} and \cite{Szmuk1}, successive convexification, a specific implementation of the SCP method, is introduced to find minimum-fuel and minimum-time trajectories in a 6-DOF rocket landing problem. In \cite{Dueri}-\cite{Mao1}, certain details of the implementation, including choice of discretization method, constraint satisfaction between temporal nodes, convexification of non-convex constraints, and algorithm convergence properties are explored. Finally, the works of \cite{Szmuk2}-\cite{Reynolds1} introduce state-triggered constraints and address real-time, onboard implementations that may produce solutions in a fraction of a second.

\subsection{Main Contributions}

The main contributions of this work are the formulation and application of two optimal control problems (OCPs):
\begin{enumerate}
	\item An off-line method to produce accurate estimates of the optimal slew time and required energy for a reference satellite to maneuver between any two arbitrary orientations. The estimates may be used by a scheduler to produce dynamically-feasible pointing schedules for each satellite of a constellation. We call the associated problem: \textit{Minimum-Time Slew OCP}.
	\item An on-line method to produce optimal attitude trajectories that satisfy desired multi-target pointing schedules with minimum control effort. This method may be run on-board to generate trajectories for a low-level tracking controller to then follow. The corresponding problem is called: \textit{Minimum-Effort Multi-Target Pointing OCP}.
\end{enumerate}
These two contributions can be applied together to support the scheduler. The first contribution allows a constellation scheduler to make informed multi-target pointing decisions (i.e., pointing schedules) based on an accurate assessment of a spacecraft's maneuvering capabilities. The second contribution allows the spacecraft to plan feasible attitude trajectories that satisfy the desired pointing schedules generated by the scheduler. 

\subsection{Organization of Paper}

After introducing the dynamical model and attitude parameterization conventions assumed in this paper, we present two problem formulations with minimum-time slew and minimum-effort multi-target pointing objectives. We describe the process of transcribing the continuous-time trajectory optimization problems into convex, finite-dimensional parameter optimization problems. We then review state-of-the-art techniques in sequential convex programming (SCP) that are used to approximately solve the original problems. Finally, we apply the formulations to numerical examples.

\subsection{Preliminaries}

To study attitude motion, a satellite may be modeled as a gyrostat \cite{Hughes}, consisting of the platform (i.e., spacecraft bus) and actuation rotors (e.g., momentum/reaction wheels). The state vector for this system of rigid bodies may be represented as:
\begin{align}
	\mathbf{x} = [ \mathbf{q}^\top \ \ \boldsymbol{\omega}^\top \ \ \mathbf{r}^\top ]^\top = [q_{\scriptscriptstyle 1}  \ \ q_{\scriptscriptstyle 2}  \ \ q_{\scriptscriptstyle 3}  \ \ q_{\scriptscriptstyle 4} \ \ \omega_{x} \ \ \omega_{y} \ \ \omega_{z} \ \ r_1 \ \ r_2 \ \ r_3 \ \ r_4 ]^\top
\end{align}
where $\mathbf{q}$ is the unit quaternion (i.e., $\lVert \mathbf{q} \rVert_{\scriptscriptstyle 2} = 1$) that describes the orientation of the spacecraft's body fixed frame with respect to an inertial frame, $\boldsymbol{\omega}$ is the angular velocity vector of the spacecraft with components expressed about the body-fixed axes, and $\mathbf{r}$ represents the angular momentum of each spinning rotor about its axis of rotation (we assume the spacecraft has four actuation rotors). The input vector consists of the torques produced by each rotor about its axis:
\begin{align}
	\mathbf{u} = [u_1 \ \ u_2 \ \ u_3 \ \ u_4 ]^\top 
\end{align}
The differential equations describing the motion of this 11-state, 4-input system consist of the quaternion kinematics, gyrostat equation, and a single-integrator model for the rotors:
\begin{align} \label{eqn:nonlinsystem}
	\dot{\mathbf{x}} &:= \frac{d}{dt}\mathbf{x} \ = \ \mathbf{f}(\mathbf{x},\mathbf{u}) := \begin{bmatrix} \frac{1}{2} \Omega \mathbf{q} \\
	-J^{-1} \big( \boldsymbol{\omega}^\times \left( J \boldsymbol{\omega} + A_r \mathbf{r} \right) + A_r \mathbf{u} \big) \\ \mathbf{u} \end{bmatrix} 
\end{align} 
The positive definite matrix $J$ represents the mass moment of inertia of the spacecraft in the body-fixed frame, $A_r \in \mathbb{R}^{3 \times 4}$ is the actuator Jacobian where each column represents a rotor's axis of rotation with respect to the body-fixed frame, and $\Omega$ and $\boldsymbol{\omega}^\times$ are skew-symmetric matrices defined as:
\begin{align} \label{eqn:skewmats}
	\Omega := \begin{bmatrix} 
	\phantom{\text{-}}0                    & \phantom{\text{-}}\omega_{z}  &                 \text{-} \omega_{y} & \phantom{\text{-}}\omega_{x} \\   
	                \text{-} \omega_{z} & \phantom{\text{-}}0                     & \phantom{\text{-}}\omega_{x} & \phantom{\text{-}}\omega_{y} \\ 
	\phantom{\text{-}}\omega_{y} &                 \text{-}\omega_{x}  & 0                                                 & \phantom{\text{-}}\omega_{z} \\
	                 \text{-} \omega_{x} &                 \text{-}\omega_{y}  &                  \text{-} \omega_{z} & 0                              \end{bmatrix} 
	\hspace{1.0cm}
	\boldsymbol{\omega}^\times := \begin{bmatrix} 
	0                                              &                 \text{-} \omega_{z} & \phantom{\text{-}}\omega_{y} \\   
	\phantom{\text{-}}\omega_{z}  & 0                                             &                 \text{-} \omega_{x} \\ 
	                \text{-} \omega_{y}  & \phantom{\text{-}}\omega_{x} &                                              0 \end{bmatrix} 
\end{align}
Note that we use the notation $[ \ ]^\times$ to map a vector into a corresponding skew symmetric matrix.
 
In this paper, we follow the unit quaternion (or Euler Symmetric Parameters) convention used in \cite{Wertz} and \cite{Markley} for attitude parameterization, where the vector part is stacked on top of the scalar part:
\begin{align}
	\mathbf{q} := [\mathbf{q}_{v}^\top \ & \ q_{s} ]^\top := [q_{\scriptscriptstyle 1}  \ \ q_{\scriptscriptstyle 2}  \ \ q_{\scriptscriptstyle 3}  \ \ q_{\scriptscriptstyle 4}  ]^\top
\end{align}
Furthermore, we denote the quaternion conjugate and the identity quaternion, respectively, as:   
\begin{align}
	\mathbf{q}^+ &:= [\text{-}\mathbf{q}_{v}^\top \ \ q_{s} ]^\top = [\text{-}q_{\scriptscriptstyle 1} \ \ \text{-}q_{\scriptscriptstyle 2}  \ \ \text{-}q_{\scriptscriptstyle 3}  \ \ q_{\scriptscriptstyle 4}  ]^\top \\
	\mathbf{q}^{\scriptscriptstyle \text{I}} &:= [ \phantom{\text{-}}\mathbf{0}^\top \ \ \phantom{_{s}} 1 ]^\top  \hspace{0.05cm} = [\phantom{\text{- }}0 \ \ \phantom{\text{- }}0 \ \ \phantom{\text{- }}0 \ \ \phantom{\text{- }}1\phantom{\text{ }}  ]^\top
\end{align}
As defined in \cite{Markley} and used in \cite{Wie1} and \cite{Wie2}, we may measure the attitude error between a given quaternion $\mathbf{q}$ and a desired quaternion $\bar{\mathbf{q}}$ by computing the error-quaternion:
\begin{align}
	\mathbf{q}^e = \begin{bmatrix} q^e_1 \\ q^e_2 \\ q^e_3 \\ q^e_4 \end{bmatrix} &:= \bar{\mathbf{q}}^+ \mathbf{q} = \begin{bmatrix} \phantom{\text{-}}\bar{q}_4 & \phantom{\text{-}}\bar{q}_3 & \text{-}\bar{q}_2 & \text{-}\bar{q}_1 \\ \text{-}\bar{q}_3 & \phantom{\text{-}}\bar{q}_4 & \phantom{\text{-}}\bar{q}_1 & \text{-}\bar{q}_2 \\ \phantom{\text{-}}\bar{q}_2 & \text{-}\bar{q}_1 & \phantom{\text{-}}\bar{q}_4 & \text{-}\bar{q}_3 \\ \phantom{\text{-}}\bar{q}_1 & \phantom{\text{-}}\bar{q}_2 & \phantom{\text{-}}\bar{q}_3 & \phantom{\text{-}}\bar{q}_4 \end{bmatrix}  \begin{bmatrix} q_1 \\ q_2 \\ q_3 \\ q_4 \end{bmatrix} \label{eqn:errorquat}
\end{align}
We note that the Hamilton product (or any quaternion multiplication operation) between two quaternions is associative and distributive, but not commutative in general, i.e., $\bar{\mathbf{q}}^+ \mathbf{q} \neq  \mathbf{q}\bar{\mathbf{q}}^+$.

\section{Problem Formulations} \label{section:problems}

In this section we introduce two optimal control problem (OCP) formulations to be solved with sequential convex programming and applied to numerical examples involving a rotor-actuated satellite under agile-pointing operations. As stated, these two formulations have the same constraints and differ only in the objective and decision variables. We conclude this section by discussing how the objectives and constraints may be modified depending on the application.

\subsection{Minimum-Time Slew} \label{subsection:mintimeOCP}

The \textit{Minimum-Time Slew OCP} addresses the time-optimal, large-angle, rest-to-rest slew problem. The integral term in (\ref{eqn:OCPA_objective}) captures the minimum-time objective and we note that the decision variables consist of the gyrostat's state and input as well as the final time $t_f$. The gyrostat equation (\ref{eqn:OCPA_wdyn}) may be modified to include other actuators, such as magnetorquers and propulsive thrusters. We may also include modeled environmental disturbances, including moments due to gravity gradient, atmospheric drag, solar radiation pressure, and the magnetic field of Earth. Furthermore, the actuator dynamics in (\ref{eqn:OCPA_rdyn}) may use higher-fidelity models that consider, for example, brushless DC motor dynamics and rotational friction. Maximum rotor momentum and maximum rotor torque bounds are enforced by (\ref{eqn:OCPA_rmax}) - (\ref{eqn:OCPA_taumax}), respectively. Finally, the initial and final conditions are given in (\ref{eqn:OCPA_ic}) - (\ref{eqn:OCPA_fc}), where the overbar notation signifies a desired or given quantity.

\begin{align}
&\minimize_{ \mathbf{q}, \boldsymbol{\omega}, \mathbf{r}, \mathbf{u}, t_f } \hspace{0.4cm} \int_{t_0}^{t_f} 1 dt \hspace{9.3cm} \label{eqn:OCPA_objective} \\
&\text{ s.t.} \ \ \dot{\mathbf{q}}(t) \ = \frac{1}{2} \Omega(t) \mathbf{q}(t) \hspace{4.7cm} \forall \ t \in [ t_0, t_f ] \label{eqn:OCPA_qdyn} \\
&\phantom{\text{s.t. }} \ \ \dot{\boldsymbol{\omega}}(t) = -J^{-1} \big( \boldsymbol{\omega}^\times(t) \left( J \boldsymbol{\omega}(t) + A_r \mathbf{r}(t) \right) + A_r \mathbf{u}(t) \big) \hspace{0.5cm} \forall \ t \in [ t_0, t_f ] \label{eqn:OCPA_wdyn}  \\
&\phantom{\text{s.t. }} \ \ \dot{\mathbf{r}}(t) \ = \mathbf{u}(t) \hspace{5.65cm} \forall \ t \in [ t_0, t_f ] \label{eqn:OCPA_rdyn}  \\
&\phantom{\text{s.t. }} \ \ r_i(t) < r_{max} \hspace{2.9cm} \ \forall \ i = 1,\ldots,4 \hspace{0.59cm} \forall \ t \in [ t_0,t_f ] \label{eqn:OCPA_rmax} \\
&\phantom{\text{s.t. }} \ \ u_i(t) < u_{max} \hspace{2.8cm} \ \forall \ i = 1,\ldots,4 \hspace{0.59cm} \forall \ t \in [ t_0,t_f ] \label{eqn:OCPA_taumax} \\
&\phantom{\text{ s.t.}} \ \ \mathbf{q}(t_0) \ = \bar{\mathbf{q}}_{inital} \ , \hspace{0.6cm} \boldsymbol{\omega}(t_0) \ = \mathbf{0} \label{eqn:OCPA_ic} \\
&\phantom{\text{ s.t.}} \ \ \mathbf{q}(t_f) = \bar{\mathbf{q}}_{final \ },  \hspace{0.6cm} \boldsymbol{\omega}(t_f) = \mathbf{0} \label{eqn:OCPA_fc} 
\end{align}

\subsection{Minimum-Effort Multi-Target Pointing} \label{subsection:minerrorOCP}

In the following \textit{Minimum-Effort Multi-Target Pointing OCP}, we maintain the same constraints as the problem above but change the objective. Furthermore, the final time $t_f$ is no longer a decision variable but a fixed parameter. Given a schedule (i.e., sequence) of desired quaternions and angular velocities at specified times, $\{ t_k, \bar{\mathbf{q}}(t_k), \bar{\boldsymbol{\omega}}(t_k) \} \ \forall \ k \in \mathbb{K}$, where $\mathbb{K}$ is a finite set of indices, the objective in (\ref{eqn:OCPB_objective}) minimizes the error at those discrete time points while also minimizing the continuous control effort. The control effort used during the trajectory may be expressed as the energy of input signal, defined as $\int_{t_0}^{t_f} \lVert \mathbf{u}(t) \rVert_{\scriptscriptstyle 2}^{\scriptscriptstyle 2} \ dt$. We note that minimizing the last term in \ref{eqn:OCPB_objective} is equivalent to minimizing the energy of the input signal. The user-defined parameter $\gamma > 0$ weighs the angular velocity error relative to the quaternion error while $\rho > 0$ weighs the control penalty term relative to the total attitude error. 

\begin{align}
&\minimize_{ \mathbf{q}, \boldsymbol{\omega}, \mathbf{r}, \mathbf{u} } \hspace{0.4cm} \sum_{k \in \mathbb{K}} \ \ \left\{ \left\lVert \bar{\mathbf{q}}^{+}(t_k) \mathbf{q}(t_k) - \mathbf{q}^{\scriptscriptstyle \text{I}} \right\rVert_2 \ + \ \gamma \left\lVert \boldsymbol{\omega}(t_k) -  \bar{\boldsymbol{\omega}}(t_k) \right\rVert_2 \right\} \ + \ \rho \int_{t_0}^{t_f} \left\lVert \mathbf{u}(t) \right\rVert_2 dt \label{eqn:OCPB_objective} \\
&\text{ s.t.} \ \ \dot{\mathbf{q}}(t) \ = \frac{1}{2} \Omega(t) \mathbf{q}(t) \hspace{4.7cm} \forall \ t \in [ t_0, t_f ] \label{eqn:OCPB_qdyn} \\
&\phantom{\text{s.t. }} \ \ \dot{\boldsymbol{\omega}}(t) = -J^{-1} \big( \boldsymbol{\omega}^\times(t) \left( J \boldsymbol{\omega}(t) + A_r \mathbf{r}(t) \right) + A_r \mathbf{u}(t) \big) \hspace{0.5cm} \forall \ t \in [ t_0, t_f ] \label{eqn:OCPB_wdyn}  \\
&\phantom{\text{s.t. }} \ \ \dot{\mathbf{r}}(t) \ = \mathbf{u}(t) \hspace{5.65cm} \forall \ t \in [ t_0, t_f ] \label{eqn:OCPB_rdyn}  \\
&\phantom{\text{s.t. }} \ \ r_i(t) < r_{max} \hspace{2.9cm} \ \forall \ i = 1,\ldots,4 \hspace{0.59cm} \forall \ t \in [ t_0,t_f ] \label{eqn:OCPB_rmax} \\
&\phantom{\text{s.t. }} \ \ u_i(t) < u_{max} \hspace{2.8cm} \ \forall \ i = 1,\ldots,4 \hspace{0.59cm} \forall \ t \in [ t_0,t_f ] \label{eqn:OCPB_taumax} \\
&\phantom{\text{ s.t.}} \ \ \mathbf{q}(t_0) \ = \bar{\mathbf{q}}_{inital} \ , \hspace{0.6cm} \boldsymbol{\omega}(t_0) \ = \mathbf{0} \label{eqn:OCPB_ic} \\
&\phantom{\text{ s.t.}} \ \ \mathbf{q}(t_f) = \bar{\mathbf{q}}_{final \ },  \hspace{0.6cm} \boldsymbol{\omega}(t_f) = \mathbf{0} \label{eqn:OCPB_fc} 
\end{align}
We note that an alternative ``constraint formulation'' of the \textit{Minimum-Effort Multi-Target Pointing OCP} is to remove the attitude penalty terms in the objective and implement them as constraints:  
\begin{align}
	\left\lVert \bar{\mathbf{q}}^{+}(t_k) \mathbf{q}(t_k) - \mathbf{q}^{\scriptscriptstyle \text{I}} \right\rVert_2 &\leq \epsilon_{\mathbf{q}} \hspace{1.25cm} \forall \ k \in \mathbb{K}  \label{errorqmetric} \\
	 \left\lVert \boldsymbol{\omega}(t_k) -  \bar{\boldsymbol{\omega}}(t_k) \right\rVert_2 &\leq \epsilon_{\boldsymbol{\omega}} \hspace{1.25cm} \forall \ k \in \mathbb{K} \label{errorwmetric}
\end{align}
where $\epsilon_{\mathbf{q}} \geq 0$ and $\epsilon_{\boldsymbol{\omega}} \geq 0$ are user-defined error tolerances. However, if the error tolerance values are set too tight for a given attitude schedule, then the problem may be infeasible. Hence, when specific error tolerance values are not required, we may choose to solve the original ``penalty formulation'' of the problem.

Additional constraints that we may include in our problem formulations are bounds on the maximum instantaneous power drawn $P_{max}$ and maximum energy consumed $E_{max}$ by the attitude control system (i.e., all four rotors combined):
\begin{align}
&\sum_{i=1}^4 \left\vert u_i(t) \cdot \frac{1}{J_r} r_i(t) \right\vert \ < \ P_{max} \quad \hspace{2.0cm} \forall \ t \in [ t_0,t_f ]  \label{eqn:powerconstraint} \\
&\int_{t_0}^{t_f} \left\{ \sum_{i=1}^{4} \left\vert u_i(t) \cdot \frac{1}{J_r} r_i(t) \right\vert \right\} dt \ <  \ E_{max}  \label{eqn:energyconstraint}
\end{align}
where $J_r$ is the rotor inertia. Apart from the equations of motion, we note that the instantaneous power and energy constraints are non-convex due to being bilinear in the decision variables of $\mathbf{u}$ and $\mathbf{r}$. In the following section we discuss how such non-convex constraints can be approximated as convex constraints.

\section{Trajectory Optimization} \label{section:trajopt}

In this section, we first describe how a general optimal control problem (OCP) is transcribed into a convex, finite-dimensional parameter optimization problem (OPT). We then review the sequential convex programming method used in this paper. Consider the following continuous-time dynamical system:
\begin{align} \label{eqn:system}
	\dot{\mathbf{x}}(t) &:= \frac{d}{dt}\mathbf{x}(t) = \mathbf{f}(\mathbf{x}(t),\mathbf{u}(t)) \hspace{1.0cm} \forall \ t \in [ t_0,t_f ]
\end{align}
defined over the given time span, where $\mathbf{x}(t) \in \mathbb{R}^{n_x}$ is the state of the system and $\mathbf{u}(t) \in \mathbb{R}^{n_u}$ is the input to the system. 

\subsection{Time Normalization}

In \cite{Szmuk1}, a procedure is introduced to transform a free-final-time optimal control problem into a finite-dimensional parameter optimization problem by normalizing time in the dynamical system (\ref{eqn:system}) by a normalization factor $t_f$:
\begin{equation}
	\tau := \frac{t}{t_f}
\end{equation}
We treat $t_f$ as a decision variable in a free-final-time problem (e.g., minimum-time OCP)  and as a constant parameter in a fixed-final-time problem (e.g., minimum-attitude-error-and-control-effort OCP). We now express the time span of the dynamical system in terms of the normalized time: $\frac{t_0}{t_f} \leq \tau \leq 1$. Since $t = t_f\tau \Rightarrow dt = t_f d\tau \Rightarrow \frac{dt}{d\tau} = t_f$, the derivative of the scaled state with respect to normalized time is:
\begin{align} \label{eqn:normalizedsystem}
	&\mathbf{x}'(t) := \frac{d}{d\tau}\mathbf{x}(t) = \frac{dt}{d\tau}\frac{d}{dt}\mathbf{x}(t) = t_f \mathbf{f}(\mathbf{x}(t),\mathbf{u}(t)) \hspace{2.0cm} \forall \ t \in [ t_0,t_f ] \\
\intertext{Since $t$ can be expressed as a function of $\tau$ and assuming $t_0 = 0$, we can represent (\ref{eqn:normalizedsystem}) in terms of normalized time:}
	&\mathbf{x}'(t(\tau)) = \mathbf{x}'(\tau) = t_f \mathbf{f}(\mathbf{x}(\tau),\mathbf{u}(\tau)) =: \mathbf{F}(\mathbf{x}(\tau),\mathbf{u}(\tau),t_f) \hspace{0.7cm} \forall \ t \in [ 0\phantom{ \ },1] \label{eqn:finalsystem}
\end{align}
In this final expression, it is clear that the dynamical system is a function of state $\mathbf{x}$, input $\mathbf{u}$, and final time $t_f$.

\subsection{Linearization} \label{Linearization}

Assuming the dynamical system described by (\ref{eqn:finalsystem}) is non-convex but differentiable, it can be approximated as convex in $\mathbf{x}$, $\mathbf{u}$, and $t_f$ with a first-order Taylor expansion about a given trajectory:
\begin{align}
    \mathbf{x}'(\tau) &= \mathbf{F}(\mathbf{x}(\tau),\mathbf{u}(\tau),t_f) \approx  A(\tau) \mathbf{x}(\tau) + B(\tau) \mathbf{u}(\tau) + \Sigma(\tau) t_f + \mathbf{e}(\tau) \hspace{0.5cm} \forall \ t \in [ 0\phantom{ \ },1]  \label{eqn:linsystem}
\end{align}
where we denote the first-order partial derivative matrices of $ \mathbf{F}(\mathbf{x}(\tau),\mathbf{u}(\tau),t_f)$ evaluated about $\{ \bar{\mathbf{x}}(\tau),\bar{\mathbf{u}}(\tau),\bar{t}_f \}$ as
\begin{align}
    A(\tau) &:= D_{\mathbf{x}}  \mathbf{F}(\bar{\mathbf{x}}(\tau),\bar{\mathbf{u}}(\tau),\bar{t}_f) \\
    B(\tau) &:= D_{\mathbf{u}}  \mathbf{F}(\bar{\mathbf{x}}(\tau),\bar{\mathbf{u}}(\tau),\bar{t}_f) \\
    \Sigma(\tau) &:= D_{t_f}  \mathbf{F}(\bar{\mathbf{x}}(\tau),\bar{\mathbf{u}}(\tau), \bar{t}_f) = \mathbf{f}(\bar{\mathbf{x}}(\tau),\bar{\mathbf{u}}(\tau)) \\
    \intertext{with the following dynamical approximation offset term:}
    \mathbf{e}(\tau) &:= - ( A(\tau) \bar{\mathbf{x}}(\tau) +  B(\tau) \bar{\mathbf{u}}(\tau) )
\end{align}

\subsection{Discretization}

As described in \cite{Hull}, the process of converting an optimal control problem into a parameter optimization problem begins by dividing the time duration of the optimal control problem into intervals using $K$ temporal nodes. The nodes may be chosen to be equally spaced, creating $K-1$ equally-sized time intervals: 
\begin{equation} \label{eqn:Knodes}
	0 =: \tau_1 < \tau_2 < \dots < \tau_k < \ldots < \tau_{K-1} < \tau_K := 1
\end{equation}
We refer to state, input, and offset terms at each node with shorthand notation: 
\begin{align}
	\mathbf{x}_k := \mathbf{x}(\tau_k), \ \mathbf{u}_k &:= \mathbf{u}(\tau_k), \ \mathbf{e}_k := \mathbf{e}(\tau_k) \hspace{1.0cm} \forall \ k = 1, \ldots, K
\intertext{For use in the following section, we collectively refer to the decision variables that we have influence over as:}
 \mathbf{z}_k &:= [\mathbf{x}^\top_k, \mathbf{u}^\top_k, t_f]^\top \hspace{1.75cm} \forall \ k = 1, \ldots, K \label{eqn:zk}
 \end{align}
We use the First-Order Hold (FOH)-interpolation based discretization method in our implementation. As demonstrated in \cite{Malyuta0}, the FOH discretization provides fast computational time and achieves similar accuracy when compared to more advanced pseudospectral methods. Furthermore, it was shown that if convex input constraints are satisfied at the nodes, then inter-nodal convex input constraint satisfaction is also guaranteed \cite{Dueri}. For convenience, we review the process used in \cite{Szmuk0}-\cite{Szmuk5} below.

The FOH interpolation represents the input within each of the $K-1$ intervals as:
\begin{equation}
	\hspace{4.0cm} \mathbf{u}({\tau}) = \lambda_k^{-} \mathbf{u}_k + \lambda_k^{+} \mathbf{u}_{k+1} \hspace{2.0cm} \forall \ \tau \in [\tau_k, \tau_{k+1}] \ , \ k = 1, \ldots, K-1
\end{equation}
where
\begin{equation}
	\lambda_k^{-} := \frac{\tau_{k+1}-\tau}{\tau_{k+1} - \tau_k} \ , \hspace{1.0cm} \lambda_k^{+} := \frac{\tau-\tau_k}{\tau_{k+1} - \tau_k}
\end{equation}
The exact discretization of (\ref{eqn:linsystem}) is then:
\begin{equation} \label{eqn:discreteeqn}
	\mathbf{x}_{k+1} = A_{k}\mathbf{x}_{k} + B_{k}^{-} \mathbf{u}_{k} + B_{k}^{+} \mathbf{u}_{k+1} + \Sigma_{k} t_f + \mathbf{e}_{k} \hspace{0.8cm} \forall \ k = 1,\ldots, K-1
\end{equation}
\begin{align}
	A_k                                    &:= \Phi(\tau_{k+1},\tau_k) \\
	B_{k}^{-}                            &:= A_k \int_{\tau_k}^{\tau_{k+1}} \Phi^{-1}(\tau,\tau_k)\lambda_k^{-}(\tau)B(\tau)d\tau  \label{eqn:Bkminus}\\
	B_{k}^{+} 	    		        &:= A_k \int_{\tau_k}^{\tau_{k+1}} \Phi^{-1}(\tau,\tau_k)\lambda_k^{+}(\tau)B(\tau)d\tau  \\
	\Sigma_{k}                        &:= A_k \int_{\tau_k}^{\tau_{k+1}} \Phi^{-1}(\tau,\tau_k)\Sigma(\tau)d\tau  \\
	\mathbf{e}_{k}  &:= A_k \int_{\tau_k}^{\tau_{k+1}} \Phi^{-1}(\tau,\tau_k)\mathbf{w}(\tau)d\tau \label{eqn:omegak}
\end{align}
The state transition matrix satisfies the following differential equation and initial condition within each interval:
\begin{equation}
	\frac{d}{d\tau} \Phi(\tau, \tau_k) = A(\tau)\Phi(\tau,\tau_k), \quad \Phi(\tau_k,\tau_k) = I^{n_x} \label{eqn:STMdyn} \\
\end{equation}
In practice, the integrands of (\ref{eqn:Bkminus})-(\ref{eqn:omegak}) along with (\ref{eqn:STMdyn}) and (\ref{eqn:finalsystem}) are numerically integrated from the start to the end of each interval using a nominal trajectory $\{ \bar{\mathbf{x}}_k,\bar{\mathbf{u}}_k, \bar{t}_f, \bar{\mathbf{e}}_{k} \}$ for $k=1,\ldots,K-1$. Note that since we initialize the numerical integration for each interval with points from a nominal trajectory, rather than the terminal points found by integration in the previous temporal interval, this approach resembles a multiple-shooting discretization method, which has shown to improve convergence of the SCP algorithm by keeping solutions closer to the nominal trajectory. In contrast, a single shooting method allows approximation errors to grow in later temporal intervals \cite{Malyuta0}.

\subsection{Sequential Convex Programming Method} 

For the example in this paper, we use the Penalized Trust Region (PTR) variant of SCP described in \cite{Reynolds1}. A key difference with  Successive Convexification (SCvx) studied in \cite{Mao1} is that PTR treats trust regions as soft constraints placed in the objective whereas SCvx enforces hard trust region constraints that are updated based on a rule. An advantage of SCvx is that convergence of this method is guaranteed. However, the method employs slack variables that may cause the approximately solved problem to be far from the original problem if they take on non-zero values in the solution. Hence, a converged solution may be infeasible for the original problem. In the following subsections we describe the PTR implementation of virtual controls and trust regions, and an approach to constraint convexification. 

\subsubsection{Virtual Controls} \label{VCs}

While executing sequential convex programming on a trajectory optimization problem, the approximated convex problem may become infeasible. This \textit{artificial infeasibility} \cite{Szmuk1} is frequently encountered in the early iterations of the algorithm when the dynamics are linearized about a poor initial guess. To alleviate this issue, slack variables called \textit{virtual controls} are added to the discrete-time equations of motion: (\ref{eqn:discreteeqn})
\begin{equation} \label{eqn:discreteeqnwithvirtual}
	\mathbf{x}_{k+1} = A_{k}\mathbf{x}_{k} + B_{k}^{-} \mathbf{u}_{k} + B_{k}^{+} \mathbf{u}_{k+1} + \Sigma_{k} t_f + \mathbf{e}_{k} +  \mathbf{v}_{k} \hspace{1.0cm} \forall \ k = 1, \ldots, K-1
\end{equation}
These virtual controls act as dynamic relaxation terms that take on nonzero values when necessary to prevent dynamic infeasibility. In turn, use of these slack variables is heavily penalized with a term in the objective:
\begin{equation} \label{eqn:Jvc}
	J_{\scriptscriptstyle vc} = w_{\scriptscriptstyle vc} \sum_{k=1}^{K} \lVert \mathbf{v}_k \rVert_{1}
\end{equation}
where $w_{\scriptscriptstyle vc}$ is a large positive weight. Minimization of the 1-norm term encourages sparsity in the virtual control vector. 

\subsubsection{Trust Regions}  \label{TRs}

To ensure that the solver does not stray too far from a nominal trajectory where the linearized model becomes less accurate, we implement a trust region cost term. In PTR, the deviation of decision variables from the solution of the previous iteration is penalized with the 2-norm of weighted deviations:
\begin{equation} \label{eqn:Jtr}
	J_{\scriptscriptstyle tr} = \sum_{k=1}^{K} \lVert W_{\scriptscriptstyle tr} (\mathbf{z}_k - \bar{\mathbf{z}}_k) \rVert_2
\end{equation}
Assuming the system has been scaled, the weight matrix $W_{\scriptscriptstyle tr}$ may be designed as a diagonal matrix with $w_{\scriptscriptstyle tr} > 0$ :
\begin{equation} \label{eqn:wtr}
	W_{tr}:= w_{tr}\cdot I^{K(n_x +n_u + 1)}
\end{equation}

\subsubsection{Constraint Convexification} \label{ConstraintConvexification}

Let us consider a general (non-convex) constraint on the continuous-time variables:
\begin{equation} \label{eqn:linearizeconstraint}
	\mathbf{g}(\mathbf{x}(\tau),\mathbf{u}(\tau),t_f) \leq 0 \hspace{0.7cm} \forall \ \tau \in [ 0\phantom{ \ },1]
\end{equation}
This constraint can be approximated by linearizing it about a nominal trajectory:
\begin{align}
	&\tilde{\mathbf{g}}(\mathbf{x}(\tau),\mathbf{u}(\tau),t_f) := 
	\mathbf{g}(\bar{\mathbf{x}}(\tau),\bar{\mathbf{u}}(\tau),\bar{t}_f) + D \mathbf{g}(\bar{\mathbf{x}}(\tau),\bar{\mathbf{u}}(\tau),\bar{t}_f) \left( \begin{bmatrix} \mathbf{x}(\tau) \\ \mathbf{u}(\tau) \\ t_f \end{bmatrix}  -\begin{bmatrix} \bar{\mathbf{x}}(\tau) \\ \bar{\mathbf{u}}(\tau) \\ \bar{t}_f \end{bmatrix}  \right) \leq 0 \hspace{0.7cm} \forall \ \tau \in [ 0\phantom{ \ },1]
\end{align}
We further approximate the constraint by explicitly enforcing it only at the time nodes of our discretization: 
\begin{align}
	\tilde{\mathbf{g}}(\mathbf{x}_k,\mathbf{u}_k,t_f) = \tilde{\mathbf{g}}(\mathbf{z}_k) \leq 0 \hspace{0.7cm} \forall \ k = 1, \ldots, K
\end{align}
Inter-nodal constraint satisfaction is not guaranteed. However, the constraints at the nodes can be carefully designed so that constraints are enforced for all time \cite{Dueri}. In a similar fashion, any non-convex cost terms may also be approximated with a Taylor series expansion about a nominal trajectory.

We may also approximate constraints involving integral terms. For example:
\begin{align}
	\int_{t_0}^{t_f} h(\mathbf{x}(t),\mathbf{u}(t)) dt \leq 0 \ \Longrightarrow \ t_f \int_{0}^{1} h(\mathbf{x}(\tau),\mathbf{u}(\tau)) d\tau \leq 0 \ \Longrightarrow \ \frac{1}{K-1} \sum_{k=1}^{K} t_f h(\mathbf{x}_k,\mathbf{u}_k) \leq 0
\end{align}
where the nonlinear term to be summed in the last inequality may also be linearized about the nominal trajectory. 

\subsubsection{Transcribed Convex Program}

The \textit{Minimum-Time OCP}, described by equations (\ref{eqn:OCPA_objective}) - (\ref{eqn:OCPA_fc}), may be transcribed into the following convex, finite-dimensional parameter optimization problem that we call \textit{Minimum-Time Slew OPT}:
  \begin{align}
  &\minimize_{ \left\{ \mathbf{z}_k, \mathbf{v}_k \right\}_{\scriptscriptstyle k=1}^{\scriptscriptstyle K}} \hspace{0.5cm}  t_f + \sum_{k=1}^{K} \left\{ w_{\scriptscriptstyle vc}\lVert \mathbf{v}_k \rVert_1 +  \lVert W_{\scriptscriptstyle tr} (\mathbf{z}_k - \bar{\mathbf{z}}_k) \rVert_2 \right\}  \\
  &\quad \text{s.t.} \nonumber \\
  & \ \phantom{\text{s.t.}}  \ \ \ \mathbf{x}_{k+1} = A_{k}\mathbf{x}_{k} + B_{k}^{-} \mathbf{u}_{k} + B_{k}^{+} \mathbf{u}_{k+1} + \Sigma_{k} T + \mathbf{w}_{k} + \mathbf{v}_{k} \hspace{1.0cm} \forall \ k = 1, \ldots, K-1 \\
  & \phantom{\text{s.t. }} \ \ \vert \mathbf{x}_{k} \vert < \mathbf{x}_{max} \ \hspace{6.05cm} \forall \ k = 1, \ldots, K \\
  & \phantom{\text{s.t. }} \ \ \vert \mathbf{u}_{k} \vert < \mathbf{u}_{max} \ \hspace{6.05cm} \forall \ k = 1, \ldots, K \\
  & \ \phantom{\text{ s.t.}} \ \ \mathbf{x}_1 \ = \bar{\mathbf{x}}_{initial}  \\
  & \ \phantom{\text{ s.t.}} \ \ \mathbf{x}_K = \bar{\mathbf{x}}_{final \ }
\end{align}
  \\
\noindent A similar process is used to transcribe equations (\ref{eqn:OCPB_objective}) - (\ref{eqn:OCPB_fc}) into the \textit{Minimum-Effort Multi-Target Pointing OPT}:
  \begin{align}
  &\minimize_{ \left\{ \mathbf{x}_k, \mathbf{u}_k, \mathbf{v}_k \right\}_{\scriptscriptstyle k=1}^{\scriptscriptstyle K}} \hspace{0.1cm}  \sum_{k \in \mathbb{K}} \ \ \left\{ \left\lVert \bar{\mathbf{q}}_k^{+} \mathbf{q}_k - \mathbf{q}^{\scriptscriptstyle \text{I}} \right\rVert_2 \ + \ \gamma \left\lVert \boldsymbol{\omega}_k -  \bar{\boldsymbol{\omega}}_k \right\rVert_2 \right\} \ + \ \sum_{k=1}^{K} \left\{ \rho \left\lVert \mathbf{u}_k \right\rVert_2 + w_{\scriptscriptstyle vc} \lVert \mathbf{v}_k \rVert_1 + \left\lVert W_{\scriptscriptstyle tr} \left(\begin{bmatrix} \mathbf{x}_k \\ \mathbf{u}_k \end{bmatrix} - \begin{bmatrix} \bar{\mathbf{x}}_k \\ \bar{\mathbf{u}}_k \end{bmatrix} \right) \right\rVert_2 \right\}  \label{eqn:OPTB_objective} \\
  &\quad \text{s.t.} \nonumber \\
  & \ \phantom{\text{s.t.}} \hspace{2.2cm}  \ \ \ \mathbf{x}_{k+1} = A_{k}\mathbf{x}_{k} + B_{k}^{-} \mathbf{u}_{k} + B_{k}^{+} \mathbf{u}_{k+1} + \Sigma_{k} T + \mathbf{w}_{k} + \mathbf{v}_{k} \hspace{1.0cm} \forall \ k = 1, \ldots, K-1 \\
  & \phantom{\text{s.t. }} \hspace{2.2cm} \ \ \vert \mathbf{x}_{k} \vert < \mathbf{x}_{max} \ \hspace{6.05cm} \forall \ k = 1, \ldots, K \\
  & \phantom{\text{s.t. }} \hspace{2.2cm} \ \ \vert \mathbf{u}_{k} \vert < \mathbf{u}_{max} \ \hspace{6.05cm} \forall \ k = 1, \ldots, K \\
  & \ \phantom{\text{ s.t.}} \hspace{2.2cm} \ \ \mathbf{x}_1 \ = \bar{\mathbf{x}}_{initial}  \\
  & \ \phantom{\text{ s.t.}} \hspace{2.2cm} \ \ \mathbf{x}_K = \bar{\mathbf{x}}_{final \ }
\end{align}
where we assume that the set of observation points is a subset of the discretization points, i.e., $\mathbb{K} \subset \{1, \ldots, K \}$.

\subsubsection{Algorithm}

The sequential convex programming algorithm used in this paper is listed in Algorithm \ref{algo:SCP} where $\mathbf{S}^0$ is an initial guess at the solution. As a notation convention used in the algorithm, superscript $i$ refers to the solution at the $i^{th}$ iteration of the algorithm:
\begin{equation} \label{eqn:SCPnotation}
  \mathbf{S}^i :=  
    \begin{cases}
      \{ \bar{\mathbf{x}}_{ 1}^{ i}, \ldots, \bar{\mathbf{x}}_{ k}^{ i}, \ldots,  \bar{\mathbf{x}}_{ K}^{ i} \} , \\
      \{ \bar{\mathbf{u}}_{ 1}^{ i}, \ldots, \bar{\mathbf{u}}_{ k}^{ i}, \ldots,  \bar{\mathbf{u}}_{ K}^{ i} \} , \\
      \quad \quad \quad \bar{t}_f^{\ i}
    \end{cases}       
\end{equation}
The algorithm stops when either (1) the user-defined maximum number of SCP iterations $N_{max}$ have been excuted, or (2) the algorithm has converged on a solution, where we define the convergence as the satisfaction of two conditions:
\begin{equation} \label{eqn:convcond}
	\left( J_{vc} \leq \epsilon_{vc} \right) \  \wedge \ \left( J_{tr} \leq \epsilon_{tr} \right)
\end{equation}
where $\epsilon_{vc}$ and $\epsilon_{tr}$ are user-defined convergence tolerances. The first condition ensures that a negligible amount of virtual controls is used, indicating that the converged solution is dynamically feasible. The second condition ensures that the solution remains sufficiently close to the nominal trajectory upon which the trajectory optimization problem was formulated. Using a convergence flag, we keep track of whether the SCP subroutine has converged or not.

\begin{algorithm}[H] \label{algo:SCP}
    $\mathbf{Input \ \ \ :   } \ \mathbf{S}^0$ \\ 
    $\mathbf{Output :} \ \mathbf{S} ^i, \text{flag}$ \\
    \For{$i = 1 : N_{max}$}{
	formulate $\mathbf{OPT}^i \left( \mathbf{S}^{i-1} \right)$ \\
        $\mathbf{S}^i \leftarrow \mathbf{OPT}^i \left( \mathbf{S}^{i-1} \right)$ \\
              \If{$\mathbf{S}^i$ \ \text{converged } }{
        flag = 1 \\
        \Return
                }
        \If{$i = N_{max}$}{
        flag = 0 \\
        \Return
                }	
    }
    \caption{Sequential Convex Programming}
\end{algorithm}

\section{Numerical Examples}

In this section we describe two examples to which we may apply the problem formulations stated in Section \ref{section:problems}. We first demonstrate how the Minimum-Time Slew OCP is used to inform a constellation scheduler of the minimum time required by a reference satellite to conduct an arbitrary slew maneuver. We then assume that the constellation scheduler has provided a desired target-pointing schedule and we plan an attitude reference trajectory using the Minimum-Effort Multi-Target Pointing OCP to realize the schedule. We model our reference satellite based on the physical parameters listed in Table (\ref{tab:params}), representative of Planet's Skysat, a satellite capable of agile maneuvering and imaging with sub-meter (50cm GSD) resolution \cite{Ryswyk}.

\begin{table}[!htbp]
\centering
\begin{tabular}{@{}lllll@{}} \toprule
 Parameter & Value & Units & Description & First Mention \\ \midrule
\ \ \ $m$            & 110 & [kg] & Satellite mass     \\
\ \ \ $l \times w \times h$            & 60 $\times$ 60 $\times$ 95 & [cm] & Satellite dimensions (cuboid)      \\
\ \ \ $J$             & $\left[ \begin{smallmatrix} 8.5 & 0.0 & 0.0 \\ 0.0 & 8.5 & 0.0 \\ 0.0 & 0.0 & 6.0 \end{smallmatrix} \right] $ & [kg m$^2$] & Satellite inertia matrix & Eqn  (\ref{eqn:OCPA_wdyn})    \\ 
\ \ \ $r_{max}$      & 0.80 & [Nms] & Maximum rotor momentum & Eqn (\ref{eqn:OCPA_rmax}) \\ 
\ \ \ $u_{max}$  & 0.06  & [Nm] & Maximum rotor torque & Eqn (\ref{eqn:OCPA_taumax}) \\ 
\ \ \ $A_r$             & $\left[ \begin{smallmatrix}
\text{-}0.68 & \phantom{\text{-}}0.68 & \phantom{\text{-}}0.68 & \text{-}0.68 \\
\text{-}0.68 & \text{-}0.68 & \phantom{\text{-}}0.68 & \phantom{\text{-}}0.68 \\
\phantom{\text{-}}0.26 & \phantom{\text{-}}0.26 & \phantom{\text{-}}0.26 & \phantom{\text{-}}0.26
\end{smallmatrix} \right] $ & [ \ ] & Actuator Jacobian (4 rotors) & Eqn  (\ref{eqn:OCPA_wdyn})    \\  \bottomrule 
\end{tabular}
\caption{Reference satellite parameters}
\label{tab:params}
\end{table}

\subsection{Minimum-Time Slew OCP Example} \label{ExampleA}

A target-sensing or data-downlinking schedule prescribes a sequence of desired instrument/antenna pointing directions at specified times. To plan a schedule for each satellite in the constellation, the scheduler must consider the time it takes for the satellite to slew from one orientation to another. For example, if the distance between two consecutively scheduled targets on the Earth's surface is large, it may not be feasible for the satellite to slew from one pointing orientation to another given the time interval between points. The slew time depends not only on a satellite's orbital motion and the relative distance between targets, but also on its mass distribution and actuator constraints. To our knowledge, a closed-form expression for computing minimum slew times between any arbitrary orientations does not exist (even for a symmetric rigid body with independent three-axis control). 

\begin{figure}[htbp!]
\centering
\includegraphics[width=0.45\textwidth]{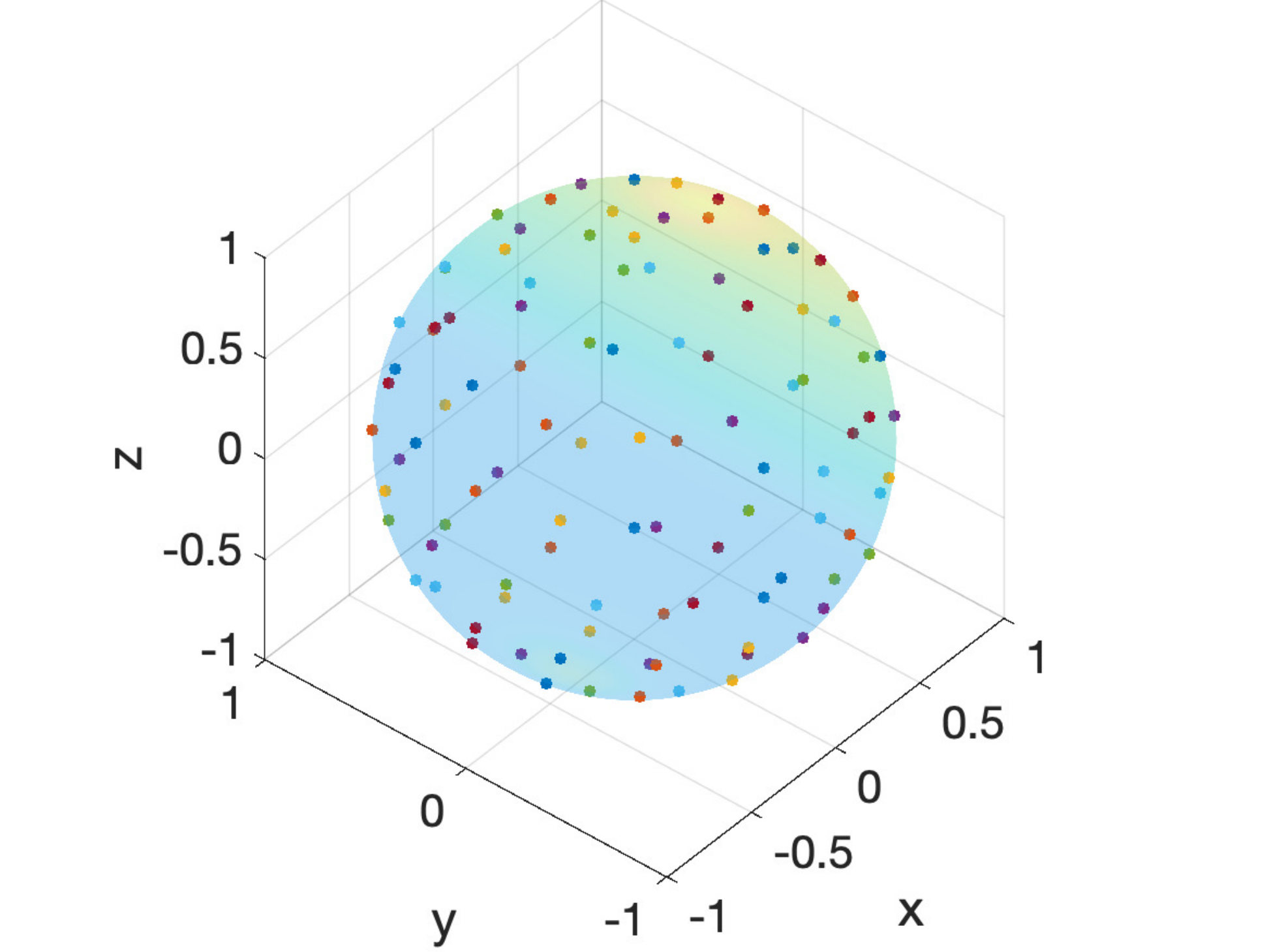}
\includegraphics[width=0.45\textwidth]{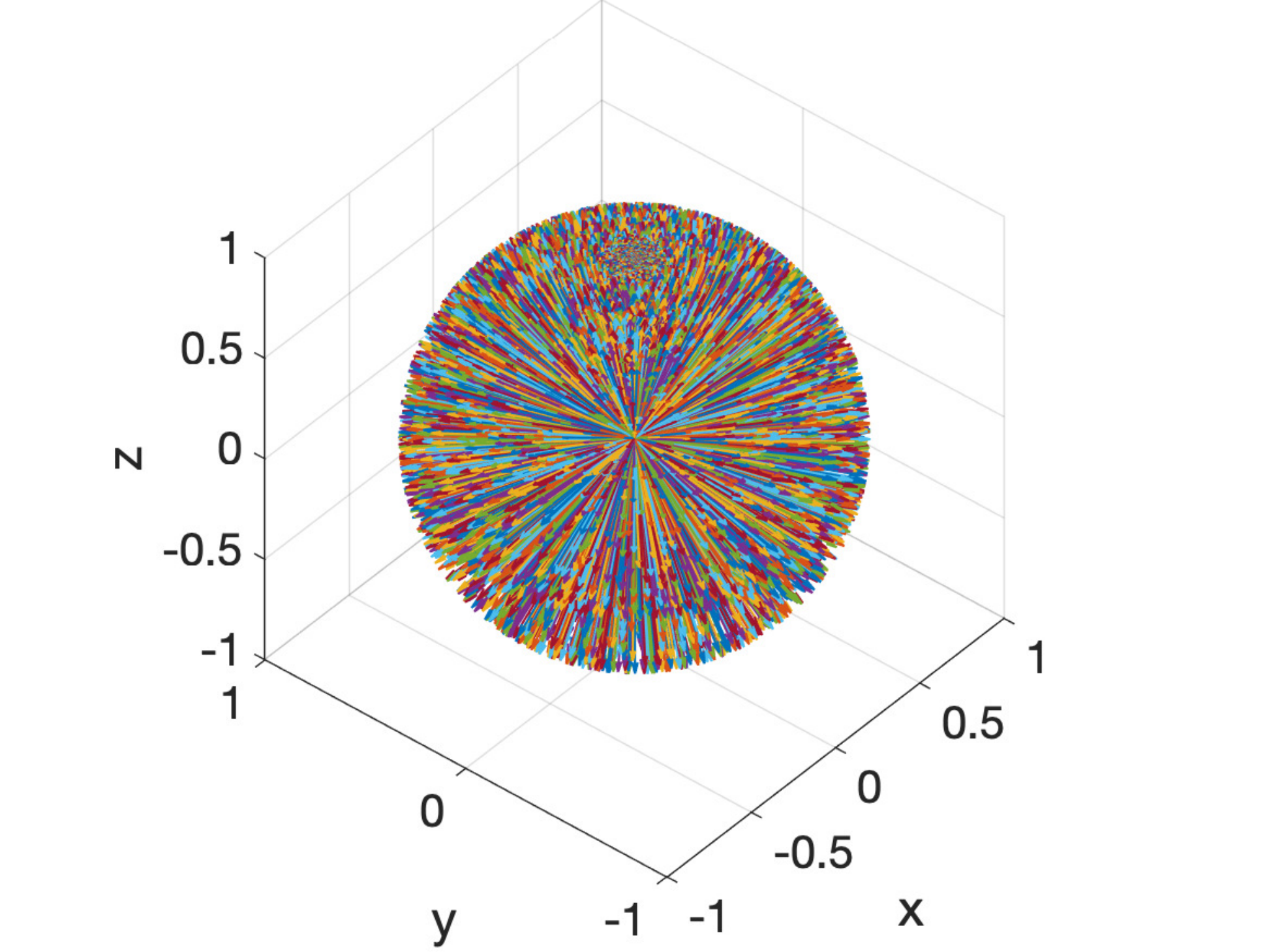}
\caption{(Left) 100 equidistributed points on unit sphere, (Right) 8,800 rotations of a pointing vector.}
\label{fig:pointingvecs}
\end{figure}

Our approach is to apply the Minimum-Time Slew OCP (Section \ref{subsection:mintimeOCP}) over a gridded space of 3D rotations. Since any attitude trajectory between any two orientations can be summarized with a single rotation about some axis, we parameterize the rotations using the Euler axis-angle representation. As illustrated in Fig. (\ref{fig:pointingvecs}), we consider 100 equidistributed axes of rotation and 88 rotations of increasing magnitude: $\theta \in \{ \text{-}180, \text{-}175, \ldots, \text{-}10, \text{-}9, \text{-}1, 1, 2, \ldots, 10, 15, \ldots, 180 \}$ degrees to cover a relatively fine grid of the entire space of 3D rotations, resulting in 8,800 unique Euler axis-angle tuples.

We set up 8,800 instances of Minimum-Time Slew OCP, where we set the initial value of the quaternion in (\ref{eqn:OCPA_ic}) to be the identity quaternion (i.e., $\mathbf{q}_0 = \mathbf{q}^{\scriptscriptstyle \text{I}}$) and the final desired quaternion in (\ref{eqn:OCPA_fc}) to be the quaternion representation of an Euler axis-angle rotation. Using the sequential convex programming method, we efficiently solve each problem instance off-line with a computation time on the order of seconds. More efficient SCP implementations \cite{Reynolds1} may solve each problem in milliseconds and the overall procedure can be parallelized for even faster off-line computation. By recording the minimum maneuver time and ADCS energy consumption from the solution to each problem instance, we can produce lookup tables or data-fitted functions to be used by a scheduler.

\subsection{Minimum-Effort Multi-Target Pointing OCP Example} \label{ExampleB}

Given a desired attitude pointing schedule $\{ t_k, \bar{\mathbf{q}}(t_k), \bar{\boldsymbol{\omega}}(t_k) \} \ \forall \ k \in \mathbb{K}$, we may plan a continuous trajectory that passes through each attitude point in the sequence by using Minimum-Effort Multi-Target Pointing OCP (Section \ref{subsection:minerrorOCP}). To motivate this method, we consider the remote-sensing application described in \cite{Nag2}, where a 24-satellite, 3-plane Walker-Delta constellation is simulated at 710 km altitude, 98.5 deg inclined, circular orbits over a 6-hour duration with the orbital mechanics module of the D-SHIELD software suite \cite{Nag4}. Simulation results include the orbital states of the satellites at each time step as well as \textit{access times} when user-defined target \textit{grid points} (described by lattitude and longitude coordinates) are observable by a satellite. In our example we consider 42 urban regions that experience frequent episodic precipitation and are prone to flooding \cite{Nag2}. Each of these globally-distributed \textit{watersheds} covers an 80 km$^2$ area spanned by 121 grid points.

\begin{figure}[htbp!]
\centering
\includegraphics[width=0.45\textwidth]{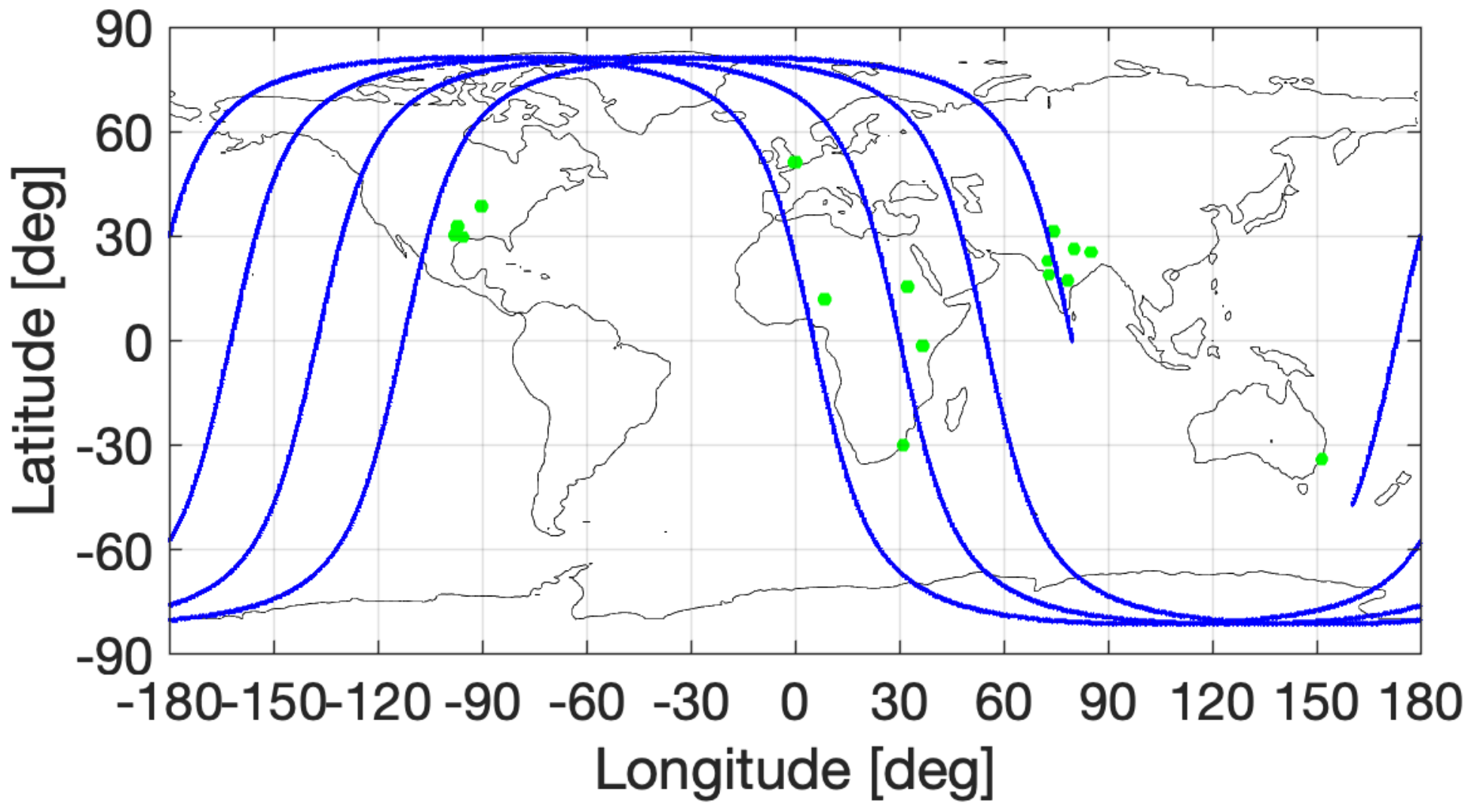}
\includegraphics[width=0.44\textwidth]{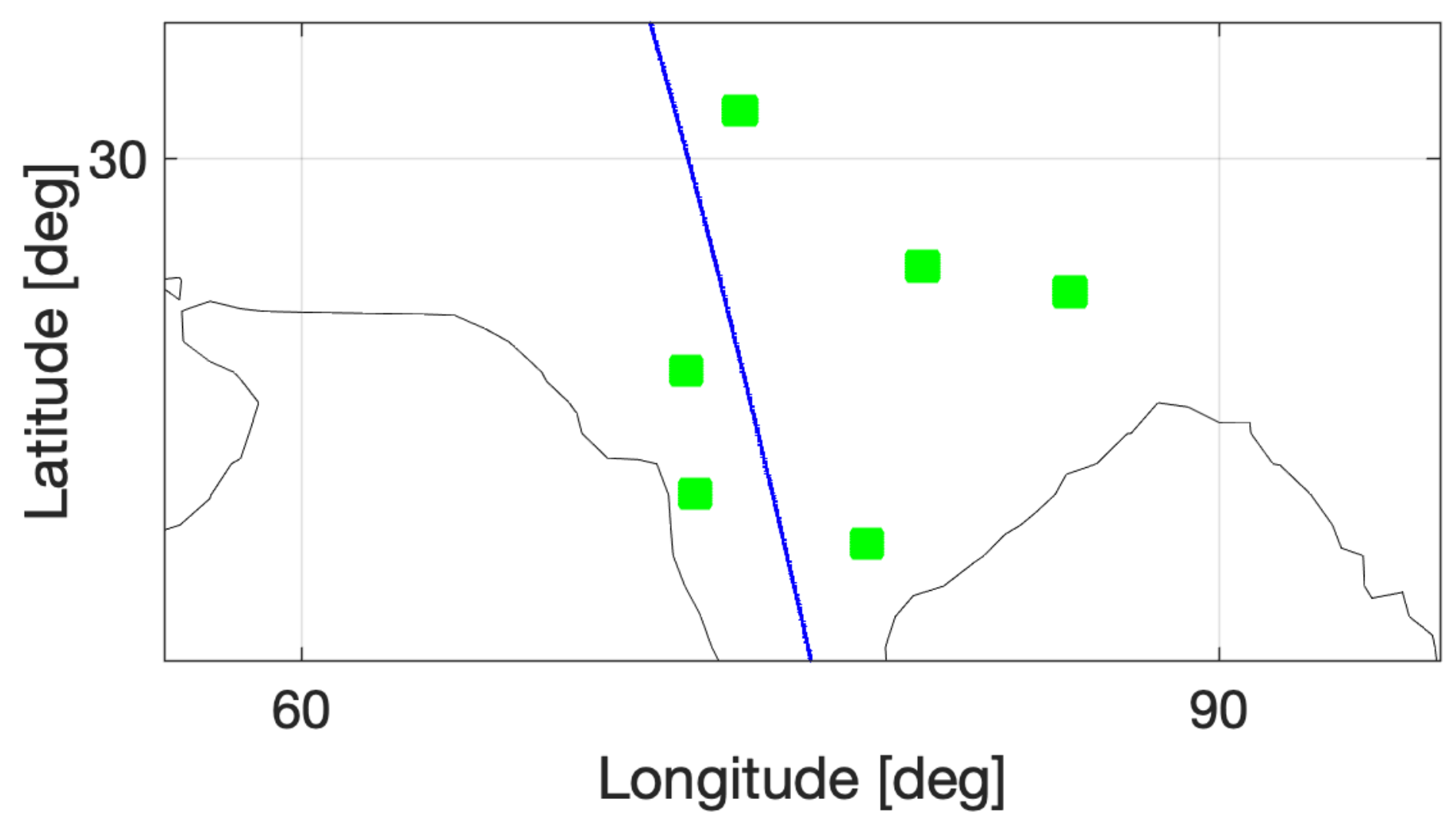}
\caption{(Left) Satellite groundtrack and 16 observable watersheds, (Right) Cluster of 6 watersheds}
\label{fig:groundtrack}
\end{figure}

In our example, we focus on a single satellite of the constellation and observe in the first plot of Fig. (\ref{fig:groundtrack}) that the satellite can access 16 of the 42 watersheds (each containing 121 green grid points) during the 6-hour time span. Only a subset of the 42 watersheds are accessible due to access restrictions, bounds on maximum off-pointing angles, or occlusion by the Earth's surface. The second plot of Fig. (\ref{fig:groundtrack}) provides a zoomed-in view of the groundtrack, containing a dense cluster of 6 regions with a total of 726 potentially accessible target grid points. As the satellite passes over this dense cluster, the satellite may be commanded by the scheduler to perform a rapid sequence of agile slewing maneuvers to acquire, point, and track desired target grid points. The commanded schedule during this part of the orbital trajectory may be the most difficult to execute by an attitude control system, requiring high-frequency intra-region slewing as well as large-angle, inter-region slewing. Hence, our example focuses on fulfilling a pointing schedule that maximizes the number of observed grid points in this 6-region cluster, treating it as a stress test for our proposed approach to executing multi-target pointing schedules. 

\begin{figure}[htbp!]
\centering
\includegraphics[width=0.45\textwidth]{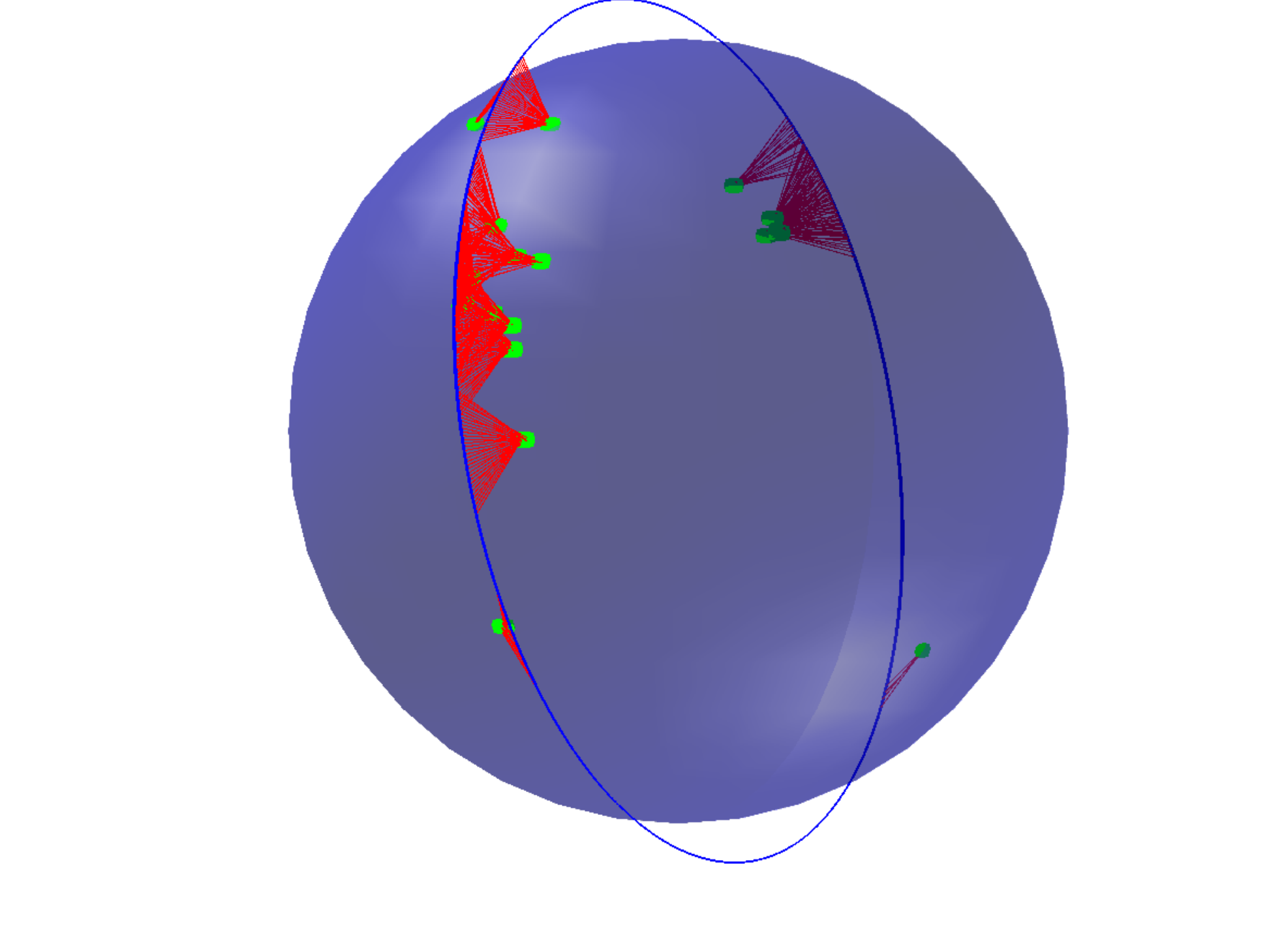}
\includegraphics[width=0.45\textwidth]{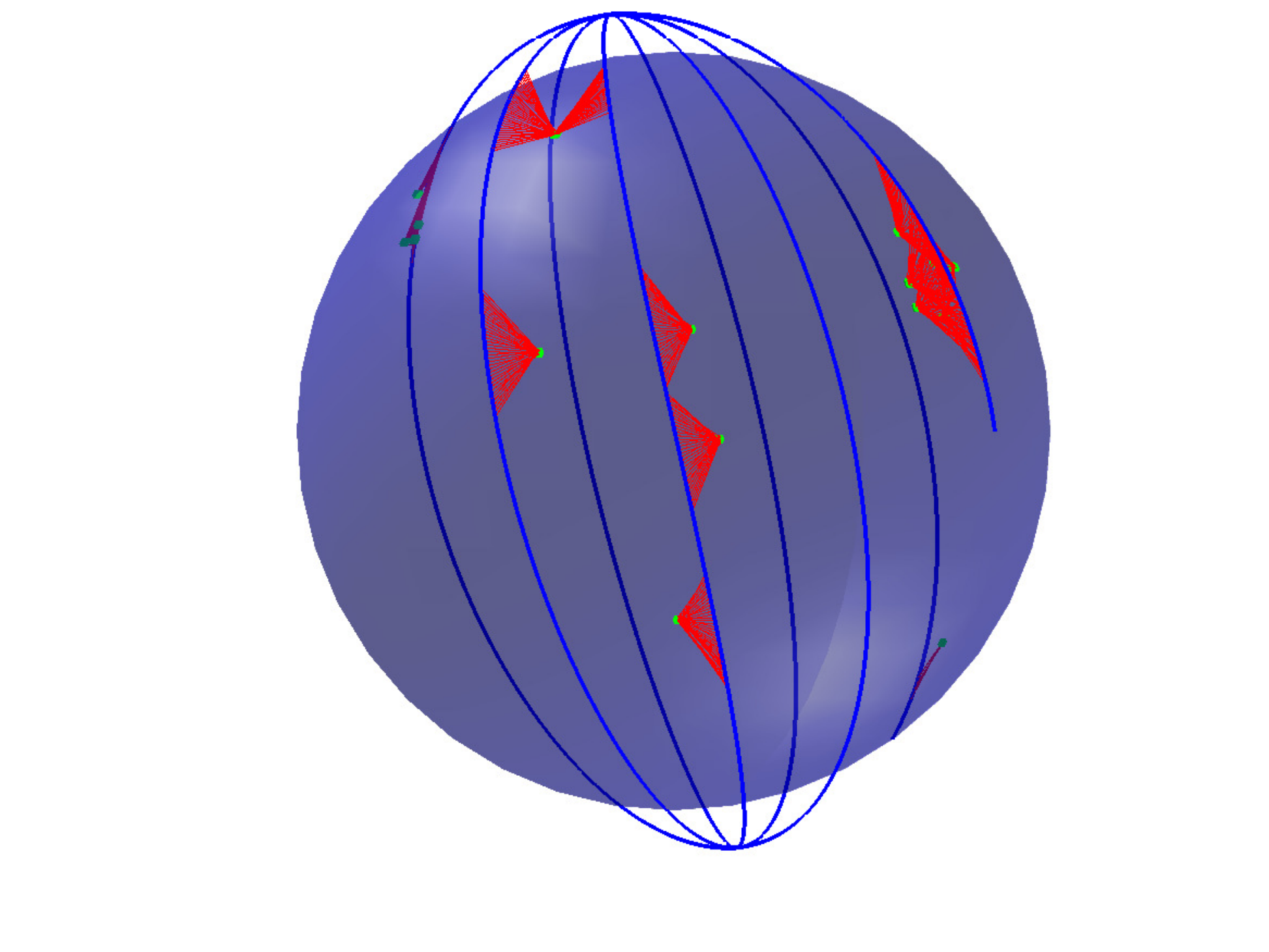}
\caption{(Left) Earth-Centered Inertial, (Right) Earth-Centered Earth-Fixed coordinate systems}
\label{fig:globe}
\end{figure}

Figure (\ref{fig:globe}) shows both Earth-Centered Inertial (ECI) and Earth-Centered Earth-Fixed (ECEF) views of the satellite's orbit, where the red vectors point from the satellite to green grid points on the Earth's surface. We note that our stress test focuses on a subset (i.e., a dense 6-region cluster) of the 16 accessible regions shown in Fig. (\ref{fig:globe}) . At any given time, a desired \textit{pointing frame} may be uniquely defined with its z-axis aligned with a red pointing vector, x-axis in the orbital plane towards the direction of motion, and y-axis completing the right-hand triad. The orientation of this frame with respect to the ECI frame is the desired pointing attitude to acquire. When a desired attitude is defined at a particular time instance, we may also define the desired body angular velocity to be zero for an interval about the specified time instance (e.g., if $\exists \ \bar{\mathbf{q}}(t_k)$, then $\{ \bar{\boldsymbol{\omega}}(t_{k-1}), \bar{\boldsymbol{\omega}}(t_k), \bar{\boldsymbol{\omega}}(t_{k+1}) \} = \mathbf{0}$). Specifying zero velocity at the time of target observation may mitigate the effects of motion blur while imaging.
	
Using the following guidelines, we manually design a sequence of desired pointing orientations (i.e., an attitude pointing schedule) to maximize grid point (GP) observations across the dense 6-region cluster:
\begin{itemize}
	\item Include as many of the 726 GPs as feasibly possible given minimum attitude slew time between observations.
	\item Visit each of the six regions (or as many as possible).  
	\item Finish observing a region before moving to the next region (i.e., cannot return to a previously observed regions).
	\item Observe GPs using a ``sweeping'' pattern, similar to the whiskbroom pattern used by early Landsat satellites \cite{Landsat}.
	\item The first accessible points in a region are observed first, deciding direction of the intra-region, sweeping pattern.
	\item The first accessible regions are observed first, deciding direction of the inter-region slewing.
\end{itemize}

As the satellite passes over the 6-region cluster, it has an approximately 10-minute window from the time when one of the 726 GPs first becomes accessible to the time that none are accessible. From an altitude of 710km, we estimate that the satellite requires at least two seconds to conduct a minimum-time, rest-to-rest slew from a grid point in the nadir direction to an adjacent grid point 8 km away. We also estimate that it can require at least 15 seconds to slew the several hundreds of kilometers between certain pairs of regions in the cluster. These estimates for slew time between neighboring GPs (intra-region slewing) and neighboring regions (inter-region slewing) are based on insights gained from Section (\ref{ExampleA}). We use the slew-time estimates to design a pointing schedule with a one-second sample time.

\begin{figure}[htbp!]
\centering
\includegraphics[width=0.45\textwidth]{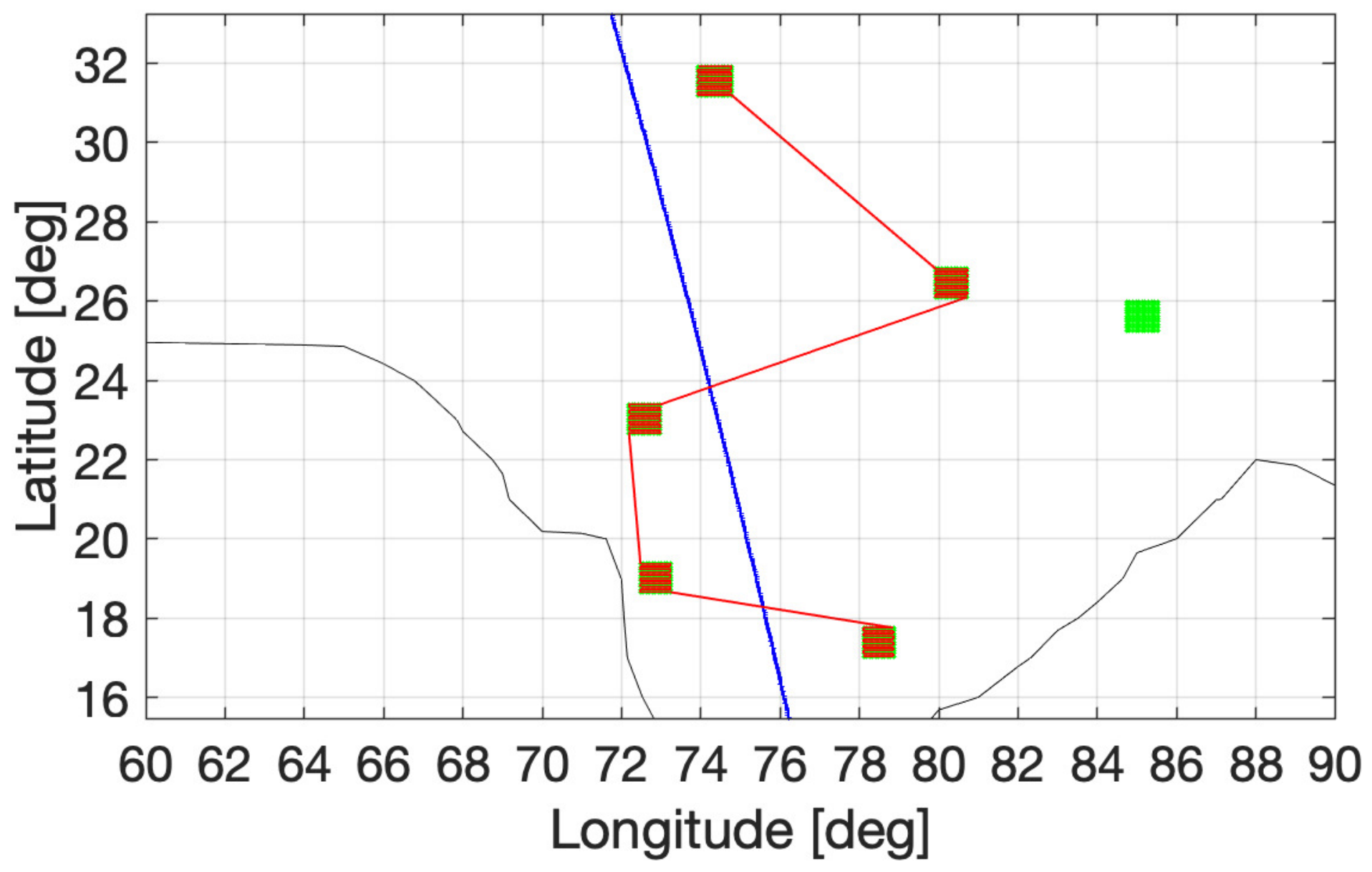}
\includegraphics[width=0.34\textwidth]{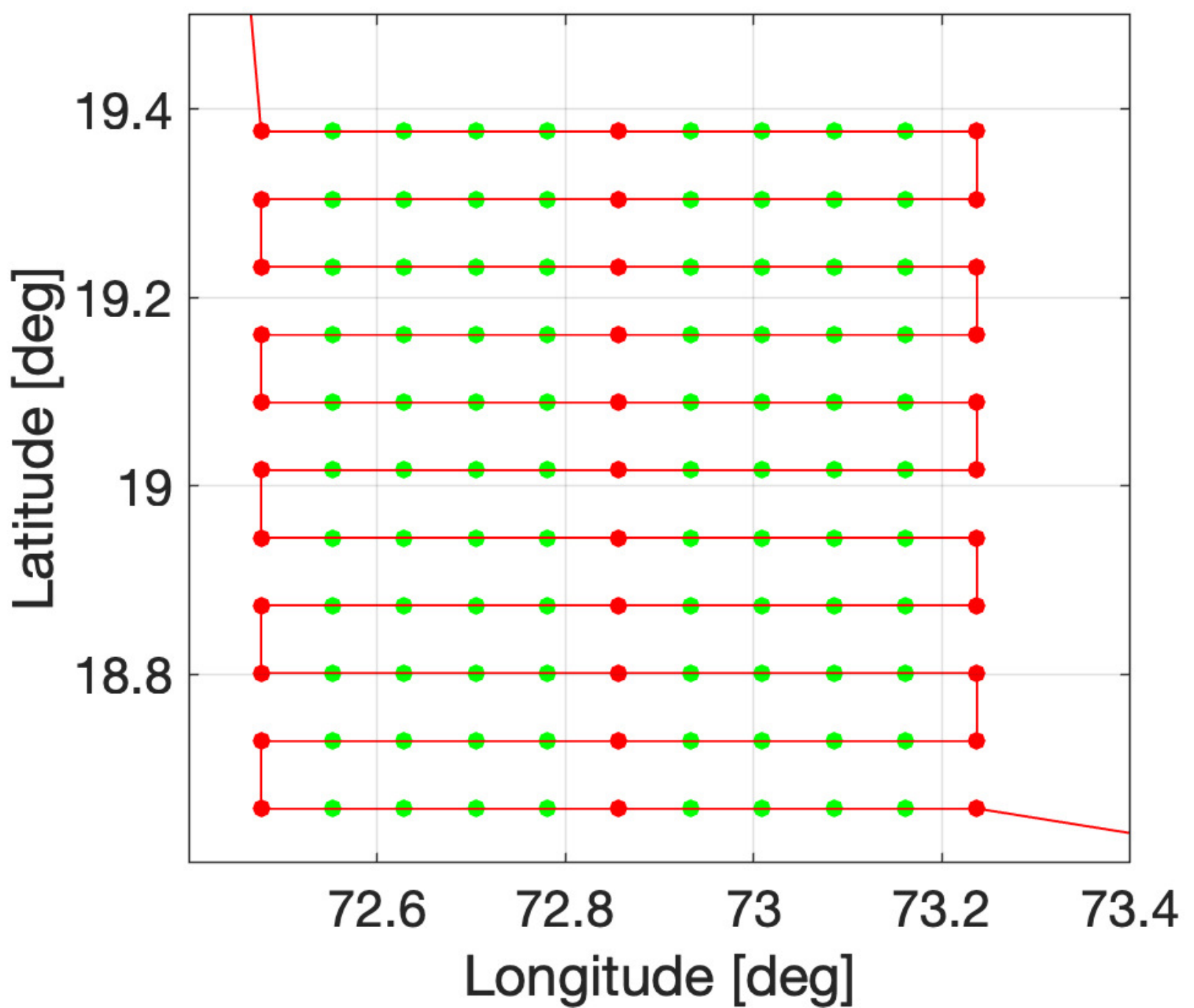}
\caption{(Left) Inter-region slewing, (Right) Intra-region slewing between 33 red GPs selected per region. }
\label{fig:schedules}
\end{figure}

The first plot of Fig. (\ref{fig:schedules}) shows a manually-designed schedule that visits five of the six regions. In the second plot of Fig. (\ref{fig:schedules}), we illustrate the ``sweeping'' pattern that we adopt within each region to observe 33 GPs per region. A red marker signifies a GP chosen for observation and a green marker is an accessible GP not chosen for observation. The red path illustrates the sequence in which the GPs are observed as well as the approximate trajectory of the satellite's pointing vector on the Earth's surface. With 33 GPs in five regions, we design a schedule that covers a total of 165 GPs or 22.7 percent of the total number of accessible GPs. In our schedule, we omit the region located near 25$^\circ$ lattitude and 85$^\circ$ longitude since its inclusion reduces the total number of GP observations, assuming the same number of GP observations per region. Given the 10-minute window and the minimum-slew time requirements between GPs and between regions, we find that the 165-GP pointing schedule is ambitious and close to the maximum number of grid points that a satellite may feasibly observe across the 6-region cluster. The other grid points must be observed on a subsequent revisit or by another satellite. Compared to this heuristics-based schedule, we expect that an optimization-based schedule could further maximize the number of GP observations. An optimization-based scheduler may explicitly use the minimum maneuver time model described in Section \ref{ExampleA} to accurately estimate the fastest transition time between GP observations. By finding the most time-efficient sequence of GPs, the scheduler can maximize the number of observations or maximize science utility \cite{Nag4}. 

\section{Results}

We apply the trajectory optimization method described Section \ref{section:trajopt} to the examples described in Sections \ref{ExampleA} and \ref{ExampleB}. For each problem, we start the SCP Algorithm (\ref{algo:SCP}) with a crude initial guess $\mathbf{S}^0$, as represented in (\ref{eqn:SCPnotation}), that consists of a spherical linear interpolated \cite{Shoemake} trajectory between desired quaternions, zero body angular velocity, zero rotor momenta, and zero rotor torque. We recall that our reference satellite is modeled after Planet's Skysat with representative physical parameters listed in Table (\ref{tab:params}). In Table (\ref{tab:SCPparams}), we list the parameters used in each problem formulation and note that the Minimum-Time Slew OPT and Minimum-Effort Multi-Target Pointing OPT use $K=$ 30 and 601 discretization nodes (\ref{eqn:Knodes}), respectively. We choose a relatively small number of nodes for the Minimum-Time Slew OPT since we expect that it is sufficient for the short time scales in this class of problems. That is, a minimum-time slew between any two orientations should be on the order of seconds. For the Minimum-Effort Multi-Target Pointing OPT, we specify a fixed final time of 600 seconds and use 601 temporal nodes so that each node represents one second of the ten-minute schedule. We then add cost terms (\ref{eqn:OPTB_objective}) for each node that corresponds to a time when a desired attitude is specified.

\begin{table}[!htbp]
\centering
\begin{tabular}{c | c c c | c c c c c | c c}
\toprule
\multicolumn{1}{c}{} & \multicolumn{3}{c}{Problem Formulation} & \multicolumn{5}{c}{SCP Algorithm}    & \multicolumn{2}{c}{Convergence Results} \\
\midrule
OPT & $K$ & $\gamma$ & $\rho$ & $N_{max}$ & $w_{vc}$ & $w_{tr}$ & $\epsilon_{vc}$ & $\epsilon_{tr}$ & $N_{converge}$ & $t_{converge}$  \\
\hline
Min-Time ($\theta=60^\circ$) &   30 &  -    &   -    & 20 & 1e5 & 1e-1 & 1e-5 & 1e-5  & 10 & 20.3  sec     \\
Min-Error, Mult-Pt & 601 &1e5 & 1e0 & 20 & 1e5 & 1e-1 & 1e-3 & 1e-4 & 9 & 144.6 sec \\
\hline
(Eqn No.) & (\ref{eqn:Knodes}) & (\ref{eqn:OPTB_objective}) & (\ref{eqn:OPTB_objective})  & Algo (\ref{algo:SCP}) & (\ref{eqn:Jvc}) & (\ref{eqn:wtr}) & (\ref{eqn:convcond}) & (\ref{eqn:convcond}) &  -    &   - \\
\bottomrule
\end{tabular}
\caption{SCP Algorithm and Objective Function parameters, Convergence results}
\label{tab:SCPparams}
\end{table}

\begin{figure}[htbp!]
\centering
\includegraphics[width=0.70\textwidth]{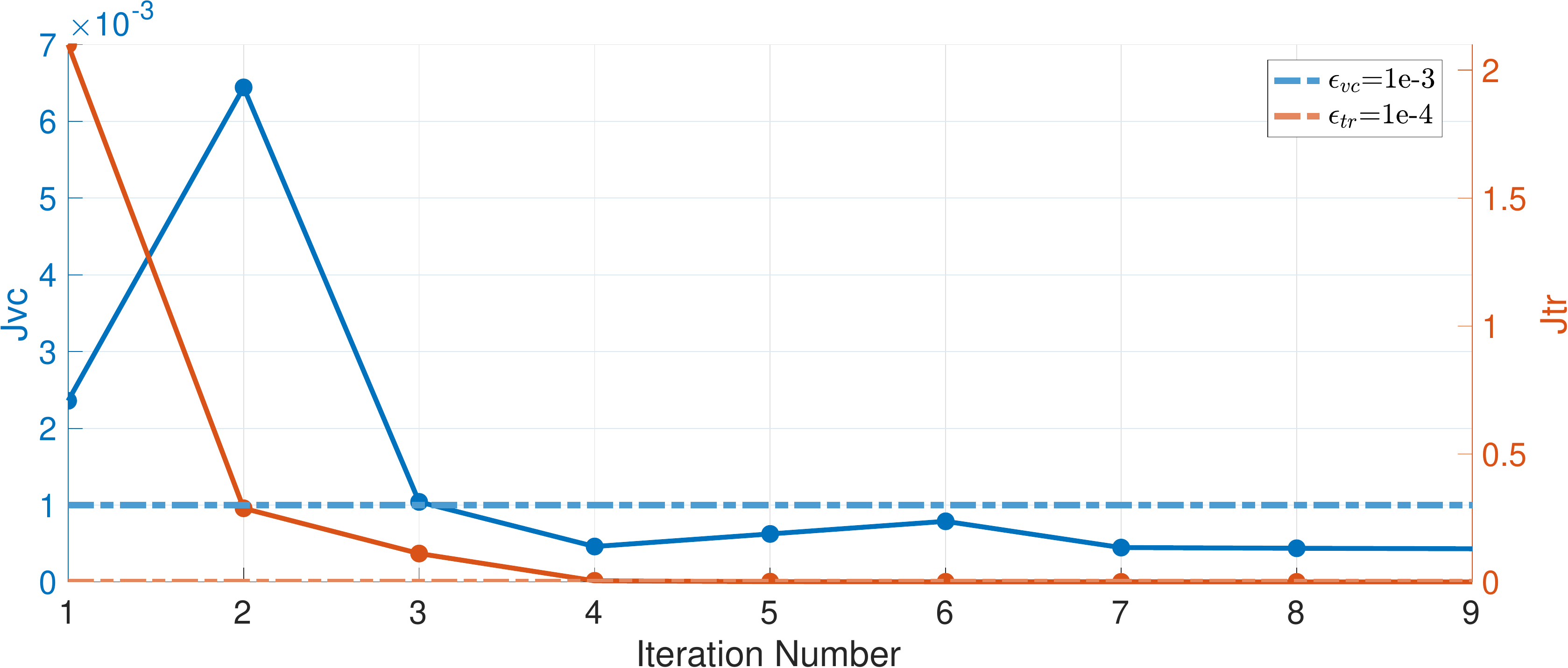} 
\caption{Virtual Control and Trust Region ($J_{vc}$, $J_{tr}$) penalty terms fall under tolerance values ($\epsilon_{vc}$, $\epsilon_{tr}$) by the 9th iteration, representing SCP algorithm convergence to a solution.}
\label{fig:cost_terms}
\end{figure}

For both OPTs, the SCP algorithm reaches convergence, as defined by equations (\ref{eqn:Jvc}) and (\ref{eqn:Jtr}), the convergence condition (\ref{eqn:convcond}), and the SCP parameters listed in Table \ref{tab:SCPparams}. Due in part to the small number of nodes used in the Minimum-Time Slew OPT, each of the 8,800 problem instances are solved in a matter of seconds. However, the Minimum-Effort Multi-Target Pointing OPT, with 601 nodes, requires 144.6 seconds to converge to a solution. Figure (\ref{fig:cost_terms}) illustrates how the virtual control and trust region penalty terms for the Minimum-Effort Multi-Target Pointing OPT decrease below user-defined tolerance values by the 9th SCP iteration. The small value for $J_{vc}$ signifies that the dynamics, state constraints and input constraints are satisfied using the input trajectory $\{\mathbf{u}(k)\}_{k=1}^K$ and only a negligible contribution from virtual controls (Section \ref{VCs}). Furthermore, the small value for $J_{tr}$ shows that the solution at the 9th iteration has not changed significantly from that of the previous iteration. This signifies that the current solution is based on an accurate linearization of the nonlinear dynamics about the previous solution, and that negligible improvement may be found in further iterations. The problems are written in MATLAB and solved using CVX, a package for specifying and solving convex programs \cite{Grant}, on a 2015 MacBook Pro with 2.5 GHz Intel Core i7 processor.

\subsection{Minimum-Time Slew OPT Result} \label{AppA}

\begin{figure}[htbp!]
\centering
\includegraphics[width=0.45\textwidth]{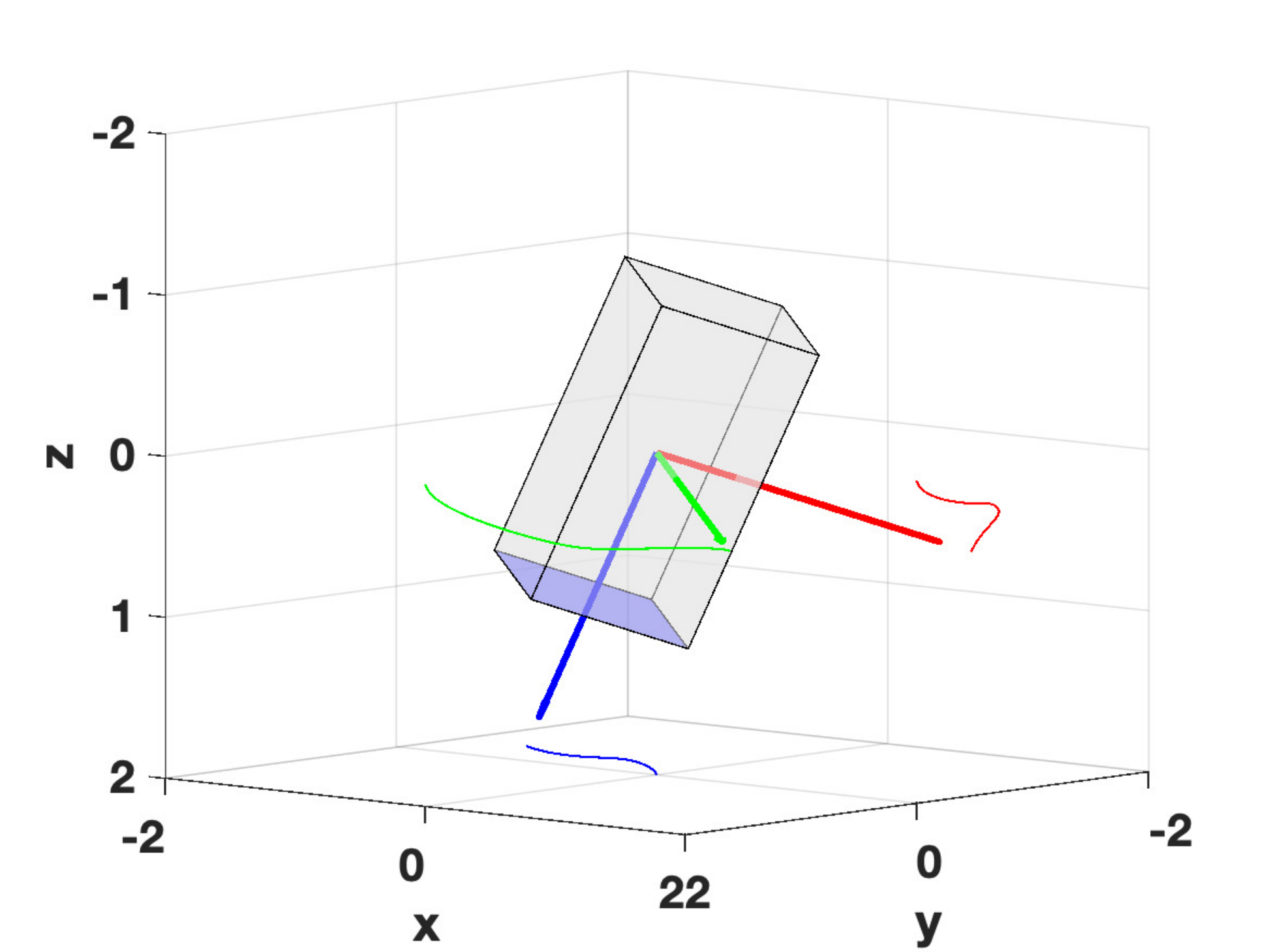} 
\caption{Minimum-time slew maneuver (60-degree rotation about an arbitrary axis)}
\label{fig:rot3Dimage}
\end{figure}

We first study the results of applying the Minimum-Time Slew OPT to a particular slew maneuver: a 60-degree rotation about an arbitrary non-principal axis. In Fig. (\ref{fig:rot3Dimage}), we illustrate the spacecraft's motion where the blue vector represents the body z-axis, aligned with an instrument's pointing vector. The trajectories traced out by the body axes represent the minimum-time slew maneuver to achieve the reorientation. We qualitatively note that this maneuver about an arbitrary non-principal axis does not resemble an eigenaxis slew. However, for slew maneuvers about any of the principal axes, we confirm that the minimum-time maneuver is an eigenaxis slew aligned with the principal axis.

We characterize the solution for this particular problem instance using five quantities: maneuver time, rotor momenta, rotor torque, ADCS instantaneous power drawn and ADCS cumulative energy consumption. The time series of these quantities are shown below. In the first plot of Fig. (\ref{fig:mintime_quaternion_angvel}), we note that the quaternion trajectory starts from a nominal orientation (represented by the identity quaternion) and reaches the desired terminal nodes, corresponding to the given axis-and-angle rotation. The second plot of Fig. (\ref{fig:mintime_quaternion_angvel}) shows that the body starts and ends with zero angular velocity, representing a rest-to-rest slew maneuver. The angular speed is greatest about the body z-axis, confirming the predominantly yaw-axis rotation depicted in Fig. (\ref{fig:rot3Dimage}). Although we have not enforced bounds on maximum body angular speed (i.e., body slew rate) in our results, we may easily add these constraints to the problem formulation. For example, to maintain star tracker accuracy we may impose an upper bound on the body slew rate. 

The optimal maneuver time is approximately 21.3 seconds. Given the same axis-and-angle and actuator constraints, we expect that a spacecraft with less mass but proportionate mass distribution requires less maneuver time. Hence, the trend towards small satellites not only results in cost efficiency but also more nimble operation.

\begin{figure}[htbp!]
\centering
\includegraphics[width=0.45\textwidth]{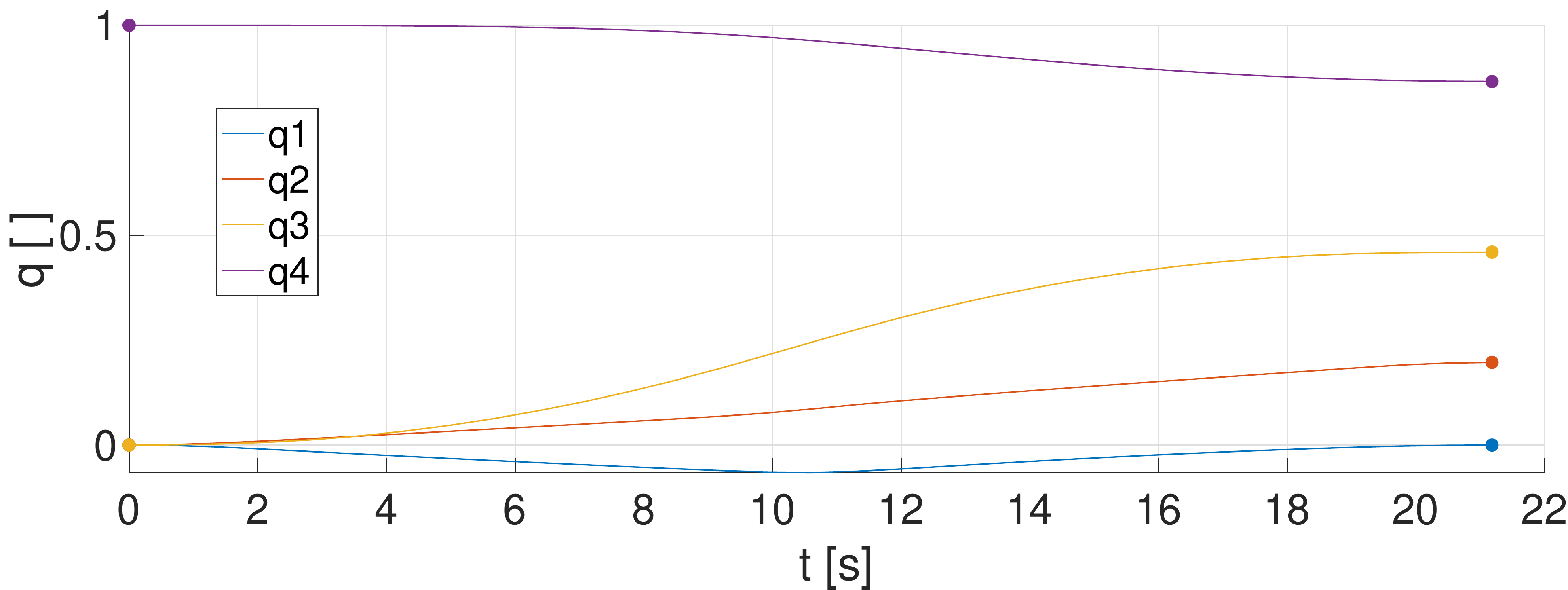} 
\includegraphics[width=0.45\textwidth]{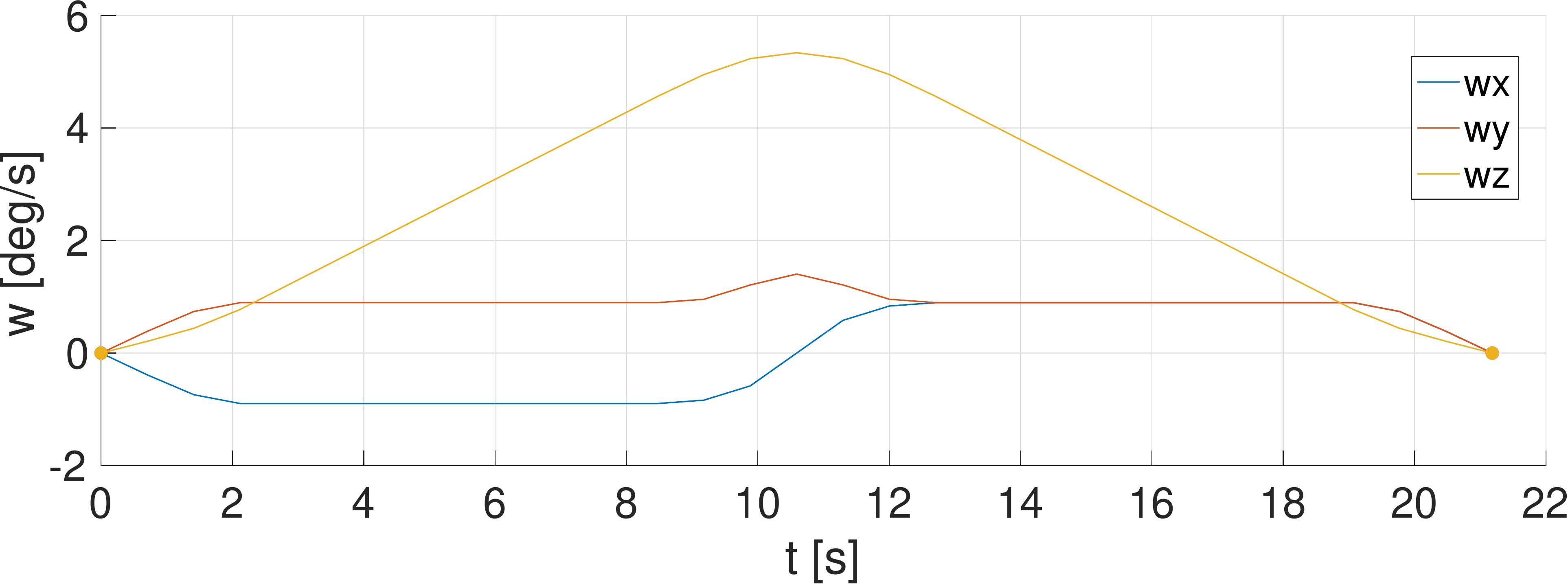}
\caption{(Left) Quaternion parameterization of attitude, (Right) Body Angular Velocity}
\label{fig:mintime_quaternion_angvel}
\end{figure}

\begin{figure}[htbp!]
\centering
\includegraphics[width=0.45\textwidth]{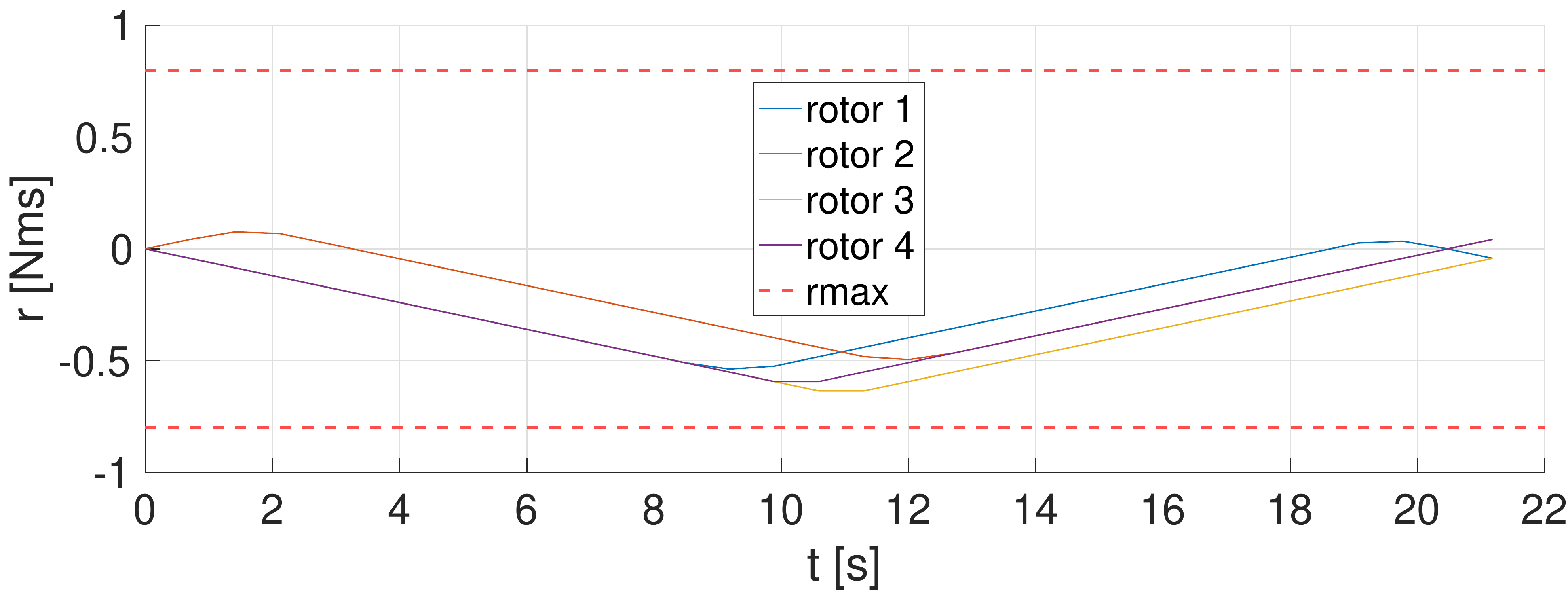}
\includegraphics[width=0.45\textwidth]{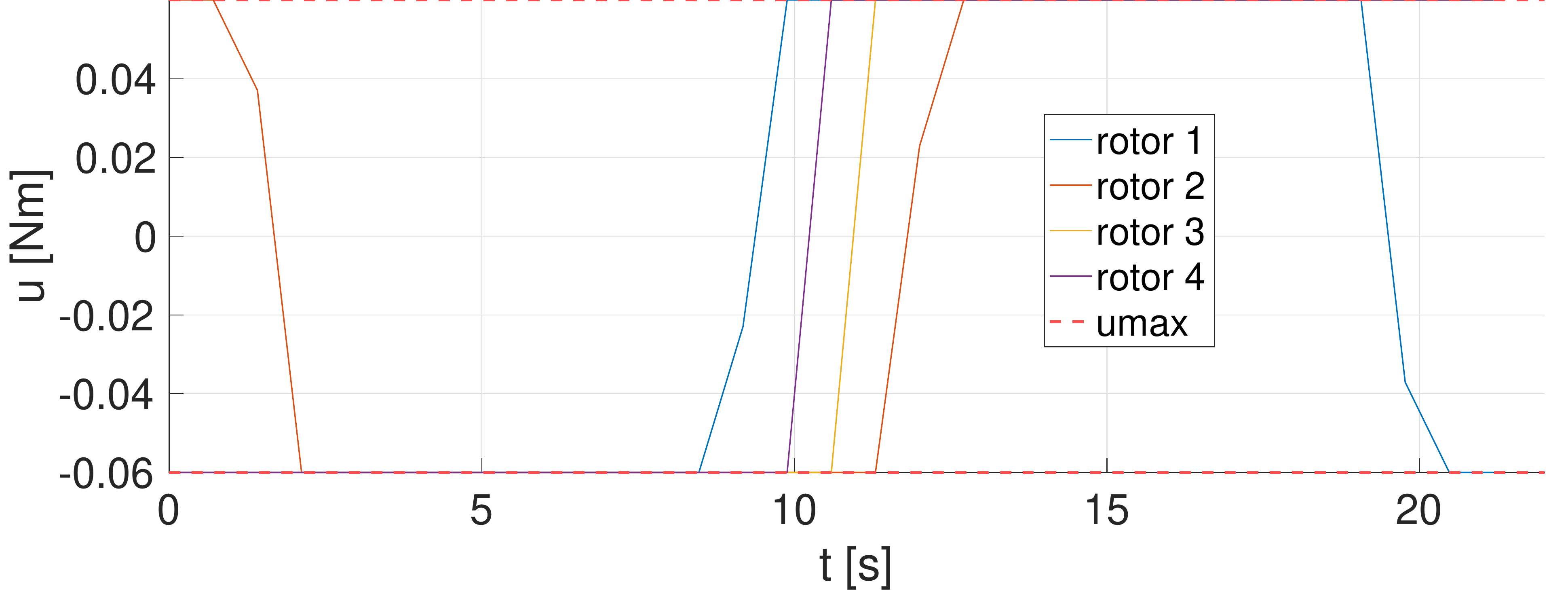}
\caption{(Left) Rotor Momentum, (Right) Rotor Torque}
\label{fig:mintime_momentum_torque}
\end{figure}

The first plot of Figure (\ref{fig:mintime_momentum_torque}) shows that the rotor momenta do not saturate for this 60-degree maneuver. For larger rotation magnitudes, we expect that the momentum trajectories may approach the magnitude bound of 0.80 Nms, impacting maneuver time. We note the non-zero rotor momenta at the end of the maneuver caused by the input torques. In practice, momentum will also build up in the rotors as they counteract torques due to environmental disturbances. This momentum buildup may be dumped with magnetorquers or other compensating mechanism. In the second plot, we observe that the rotor torque magnitudes saturate at 0.06 Nm and follow a ``bang-bang'' structure, consistent with classical minimum-time maneuvers \cite{Athans}-\cite{Bryson0}. This observation confirms our intuition that performing a rotation in minimum time requires the actuators to perform at their limits.

\begin{figure}[htbp!]
\centering
\includegraphics[width=0.45\textwidth]{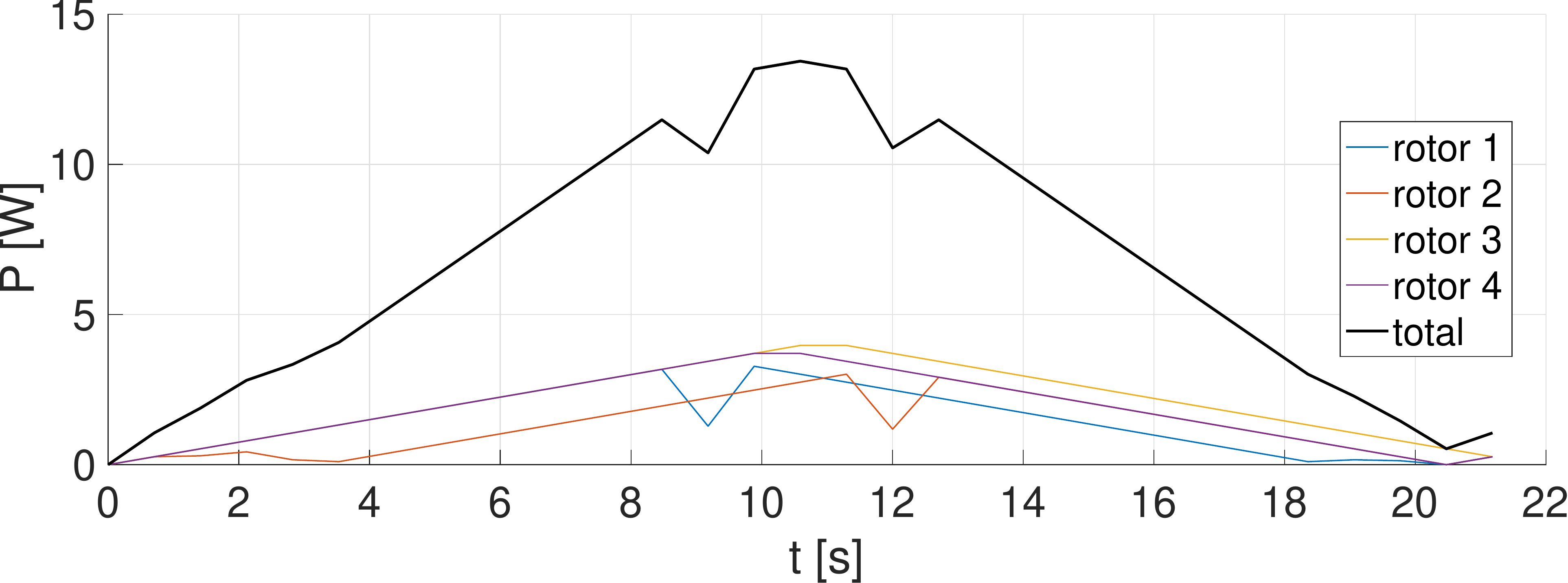}
\includegraphics[width=0.45\textwidth]{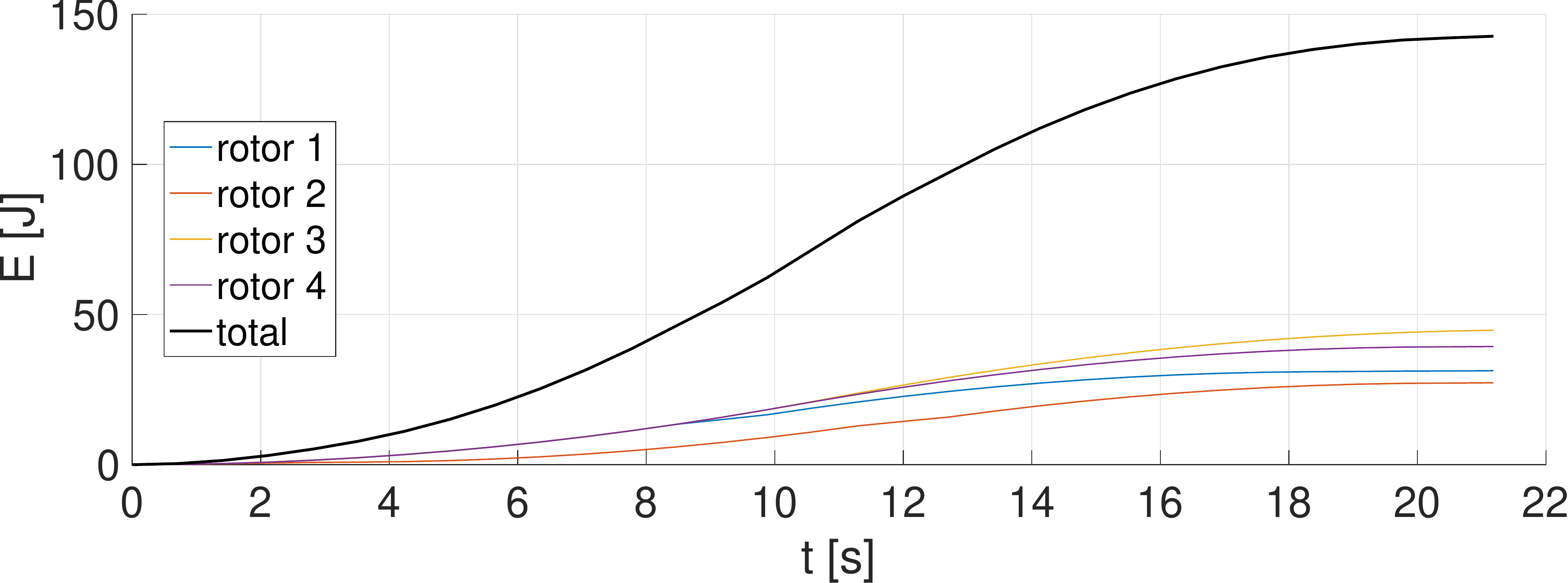}
\caption{(Left) Instantaneous Power drawn, (Right) Cumulative Energy Consumed}
\label{fig:mintime_power_energy}
\end{figure}

Finally, in Fig. (\ref{fig:mintime_power_energy}), we note that the peak instantaneous power drawn by the ADCS stays below 15 W and that the total energy consumption is less than 150 J for the maneuver. Given mission constraints on the instantaneous power or energy, we may enforce constraints (\ref{eqn:powerconstraint}) and (\ref{eqn:energyconstraint}).

\begin{figure}[htbp!]
\centering
\includegraphics[width=0.45\textwidth]{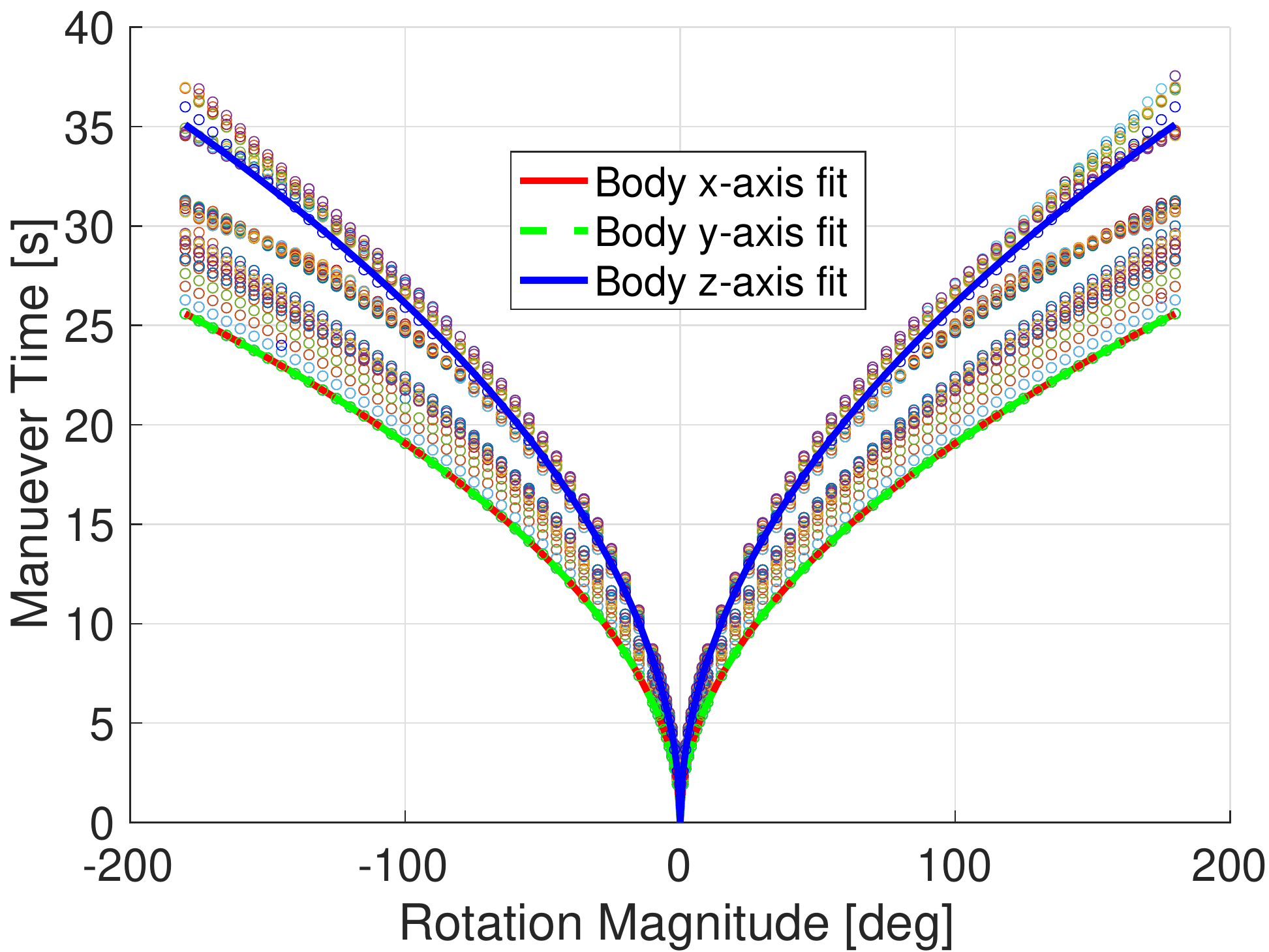}
\includegraphics[width=0.45\textwidth]{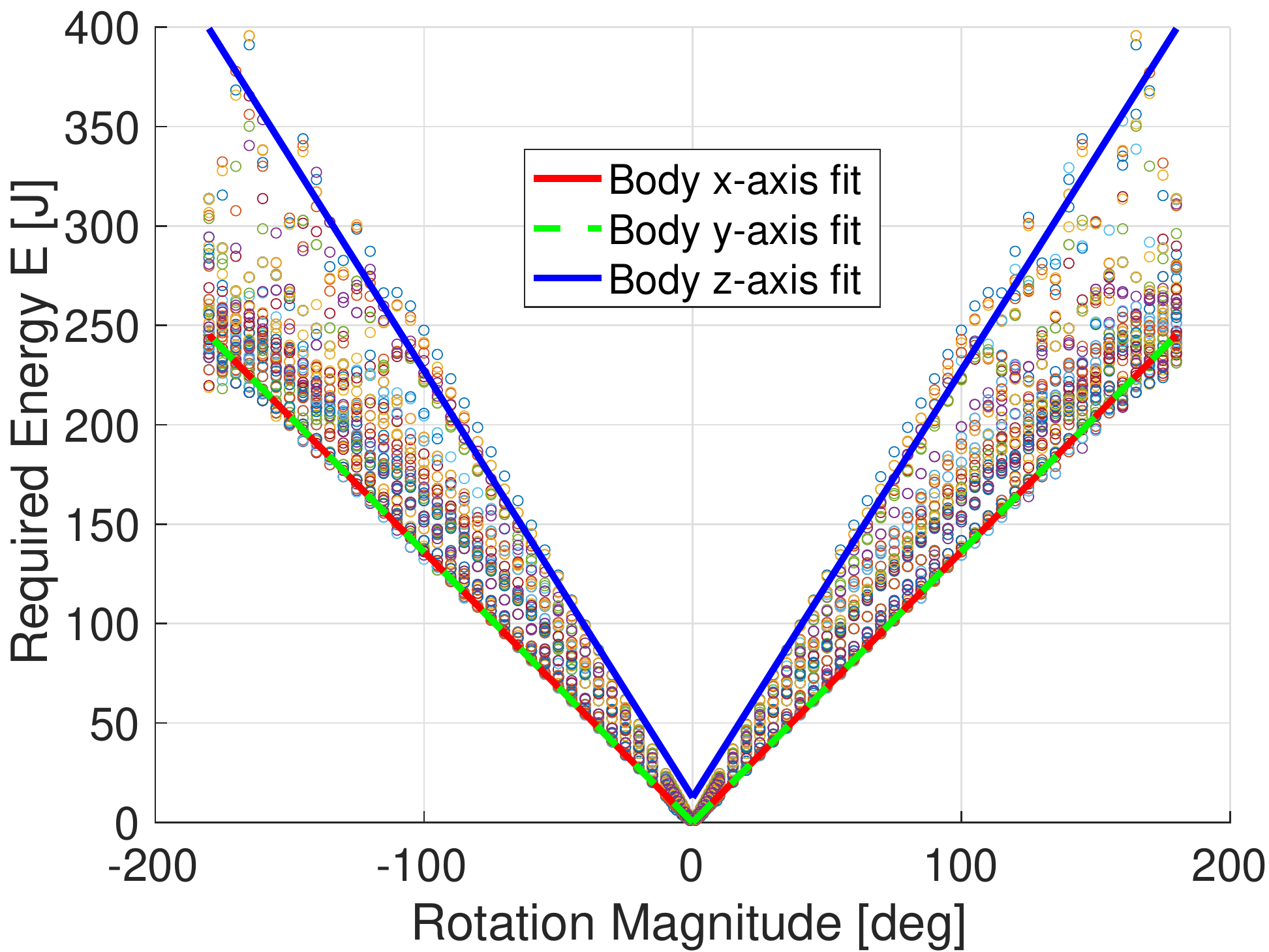}
\caption{(Left) Minimum-time, (Right) Energy as functions of eigenaxis rotation magnitude $\theta$}
\label{fig:times_energies}
\end{figure}

The results above are for one of the 8,800 problem instances described in (\ref{ExampleA}). After solving each problem instance, we plot the corresponding slew maneuver time and ADCS energy consumption as functions of the rotation magnitude in Fig. (\ref{fig:times_energies}). For each of the 100 equidistributed axes of rotation, we observe that optimal maneuver time is a power function of the rotation magnitude and that required energy is a linear function of rotation magnitude. These relationships allow us to approximate the minimum maneuver time and required energy consumption to conduct any arbitrary rotation by the reference satellite described in Table (\ref{tab:params}). The relationships may be represented in lookup tables or function approximators that can be quickly evaluated by an on-line constellation scheduler.

To empirically verify the optimality of the solutions found by our approach, we solve the Minimum-Time Slew OPT for rotations about the principal axes (i.e., about body x-, y-, and z-axes). Using least-squares fitting, we determine the coefficients for a minimum-time power model, $T=a \theta^b$, and an energy linear model, $E= c \theta + d$, valid for eigenaxis rotations $\vert \theta \vert > 1^{\circ}$ about each axis. As shown in Table (\ref{tab:coefficient}), the power coefficient in the minimum-time model approximately takes on the value of $b=0.5$ for principal axis rotations. Hence, we observe that the optimal maneuver time for a principal axis rotation is proportional to the square-root of the rotation magnitude (in radians). We note that the well-known rest-to-rest, minimum-time, double-integrator problem also reveals this relationship, as mentioned in \cite{Bilimoria} and proved in \cite{Athans} - \cite{Bryson0}. Intuitively, we may infer that the minimum-time trajectory about a principal axis is equivalent to the minimum-time eigenaxis slew about the axis, which in turn can be modeled as the double-integrator problem. Since the maneuver is purely about a principal axis, the body angular velocities are decoupled, simplifying the dynamics into a double-integrator system from torque input to rotation magnitude. This result provides assurance that our trajectory optimization method is indeed finding approximately optimal values and solutions. 

\begin{table}[!htbp]
\centering
\begin{tabular}{@{}lllll@{}} \toprule
Body Axis & \ \ \ a & \ \ \ b & \ \ \ c & \ \ \ d \\ \midrule
\ \ \ \ \ \ x            & 14.4371 & 0.5000 & 78.0329 & \text{-}0.0225    \\
\ \ \ \ \ \ y            & 14.4371 & 0.5000 & 78.0329 & \text{-}0.0225      \\
\ \ \ \ \ \ z            & 19.7292 & 0.5033 & 123.1154 & 12.5189    \\  \bottomrule 
\end{tabular}
\caption{Coefficients produced by least squares fitting (valid for $\theta > 1^{\circ}$)}
\label{tab:coefficient}
\end{table}

Beyond double-integrator dynamics about principal axes, our approach can be applied to high-fidelity spacecraft models and arbitrary rotations. Since the Minimum-Time Slew OPT can be applied to non-rest-to-rest and non-principal axis rotations, where nonlinear dynamical coupling exists, the minimum-slew-time/energy estimates found by our approach may be more accurate than those based on the classical minimum-time, rest-to-rest eigenaxis slew solution. Furthermore, we may explicitly consider system- or actuator-specific constraints, such as slew rate or rotor momentum bounds, that are not included in the classical formulation.

\subsection{Minimum-Effort Multi-Target Pointing OPT Result}  \label{AppB}

\begin{figure}[htbp!]
\centering
\includegraphics[width=0.70\textwidth]{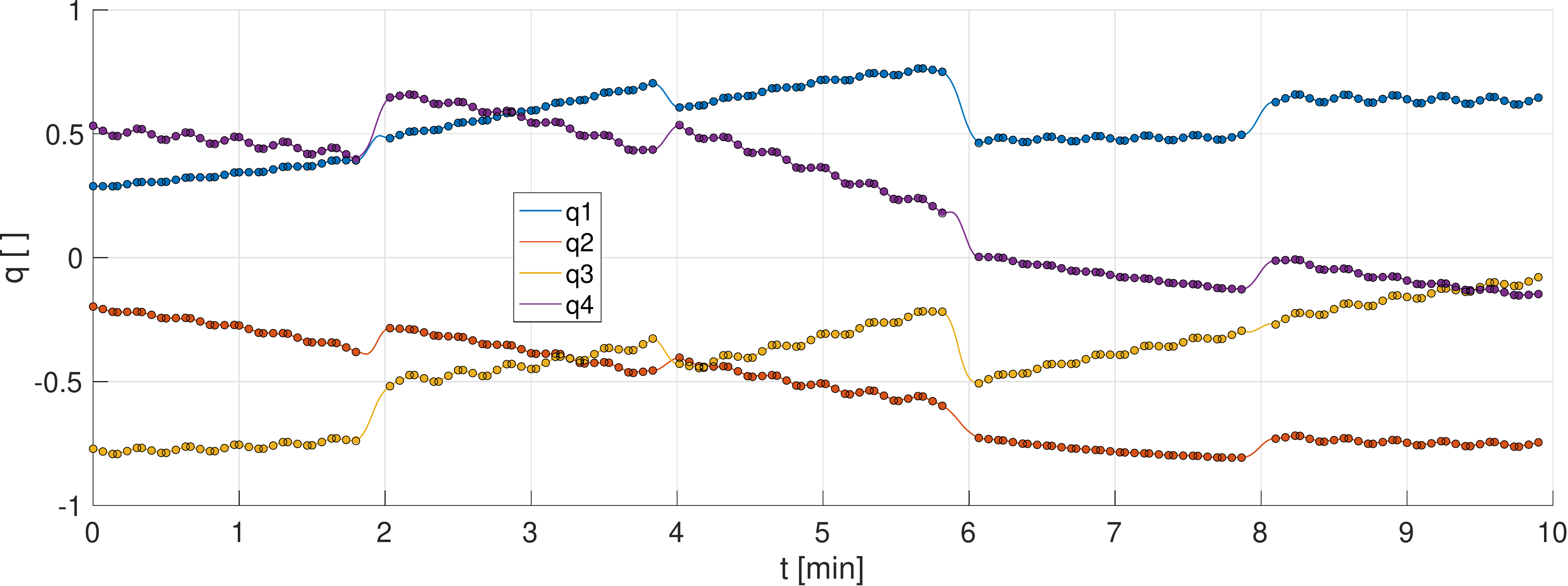}
\caption{Quaternion parameterization of attitude}
\label{fig:minerror_quaternion}
\end{figure}

\begin{figure}[htbp!]
\centering
\includegraphics[width=0.70\textwidth]{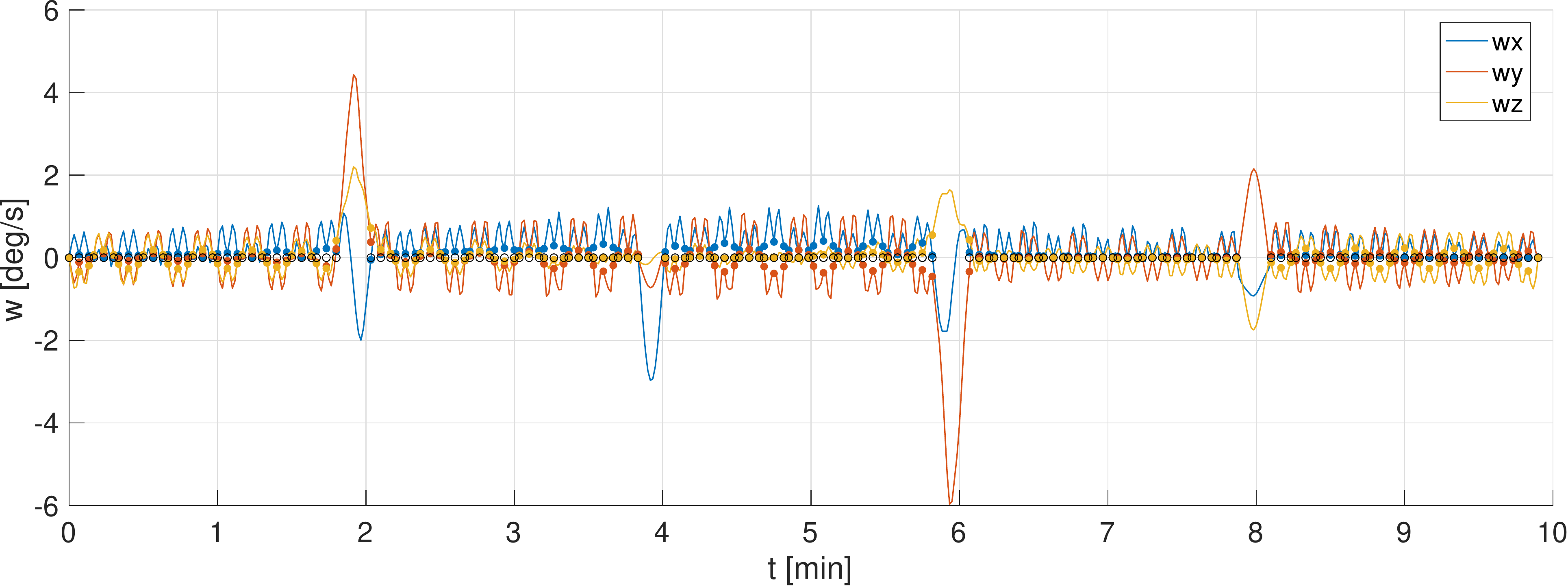}
\caption{Body Angular Velocity}
\label{fig:minerror_angvel}
\end{figure}

We apply the Minimum-Effort Multi-Target Pointing OPT to the 165-GP attitude pointing schedule for the 6-region cluster example described in Section (\ref{ExampleA}). We recall that the schedule is a discrete sequence of desired pointing orientations: $\{ t_k, \bar{\mathbf{q}}(t_k), \bar{\boldsymbol{\omega}}(t_k) \} \ \forall \ k \in \mathbb{K}$, where $\lvert \mathbb{K} \rvert = 165$. In figures (\ref{fig:minerror_quaternion}) and (\ref{fig:minerror_angvel}), we represent the 165 desired quaternion or angular velocity points with black, un-filled, circular markers. As shown in Fig. (\ref{fig:minerror_quaternion}), we find a feasible, continuous quaternion trajectory (represented by the colored lines with filled, circular markers) that passes through all desired points at the specified times. 
To quantify the error in the quaternion trajectory with respect to the desired schedule, we introduce performance metrics. Equation (\ref{eqn:qmaxmetrics}) measures the largest point-wise error between a quaternion trajectory and the desired quaternion schedule it attempts to satisfy. Equation (\ref{eqn:qavgmetrics}) determines the average error across all points. As recorded in Table \ref{tab:trajopt_vs_desired_errors}, we compute the maximum point-wise quaternion error to be $q_{e}^{max}=0.0079$ and the average quaternion error to be $q_{e}^{avg}=0.0014$.
\begin{align} 
	 q_{e}^{max} &= \max_{k \in \mathbb{K}} \ \left\lVert \bar{\mathbf{q}}_k^{+} \mathbf{q}_k - \mathbf{q}^{\scriptscriptstyle \text{I}} \right\rVert_2 \label{eqn:qmaxmetrics} \\
	 q_{e}^{avg}  &= \frac{1}{\vert \mathbb{K} \vert}  \sum_{k \in \mathbb{K}} \ \left\lVert \bar{\mathbf{q}}_k^{+} \mathbf{q}_k - \mathbf{q}^{\scriptscriptstyle \text{I}} \right\rVert_2 \label{eqn:qavgmetrics}
\end{align}

In Fig. (\ref{fig:minerror_angvel}), we show the corresponding angular velocity trajectory where we observe that the spacecraft reaches nearly-zero angular velocity at all 165 desired points. Similar to the performance metrics for the quaternion trajectory, we also introduce equations (\ref{eqn:wmaxmetrics}) and (\ref{eqn:wavgmetrics}) to quantify the maximum point-wise error and the average error in the angular velocity trajectory with respect to the desired schedule. We record a maximum point-wise error of $\omega_{e}^{max}=0.8135$ and an average error of $\omega_{e}^{avg}=0.1980 $.
\begin{align} 
	 \omega_{e}^{max} &= \max_{k \in \mathbb{K}} \ \left\lVert \boldsymbol{\omega}_k -  \bar{\boldsymbol{\omega}}_k \right\rVert_2 \label{eqn:wmaxmetrics} \\
	 \omega_{e}^{avg}  &= \frac{1}{\vert \mathbb{K} \vert}  \sum_{k \in \mathbb{K}} \ \left\lVert \boldsymbol{\omega}_k -  \bar{\boldsymbol{\omega}}_k \right\rVert_2 \label{eqn:wavgmetrics}
\end{align}
To further reduce the angular velocity error, we may increase the relative state penalty $\gamma$ in the Minimum-Effort Multi-Target Pointing OPT objective (\ref{eqn:OPTB_objective}). We may also choose to include an explicit constraint on the error using (\ref{errorwmetric}) and attempt to find a feasible solution.

We also note in Fig. (\ref{fig:minerror_angvel}) that the angular velocity components peak while slewing between the five regions. The large peaks are due to the limited time spans allotted for the inter-region slews specified in our manually-designed schedule. Imposing body slew rate constraints would reduce the amplitude of the peaks but may also result in increased quaternion error, especially for the initial points of each region when high slew rates are required to move quickly from the previous region. A ``relaxed'' schedule that allows for more time to conduct inter-region slews could satisfy slew rate constraints with minimal quaternion error but at the expense of observing more GPs.

\begin{table}[!htbp]
\centering
\begin{tabular}{c c | c c}
\toprule
\multicolumn{2}{c}{Quaternion Error}  & \multicolumn{2}{c}{Angular Velocity Error}  \\
\midrule
$q_{e}^{max}$ & $q_{e}^{avg}$ & $\omega_{e}^{max}$ & $\omega_{e}^{avg}$ \\
\hline
0.0079 & 0.0014 & 0.8135 & 0.1980   \\
\bottomrule
\end{tabular}
\caption{Reference trajectory error with respect to desired GP schedule}
\label{tab:desiredGPerror}
\end{table}

\begin{figure}[htbp!]
\centering
\includegraphics[width=0.70\textwidth]{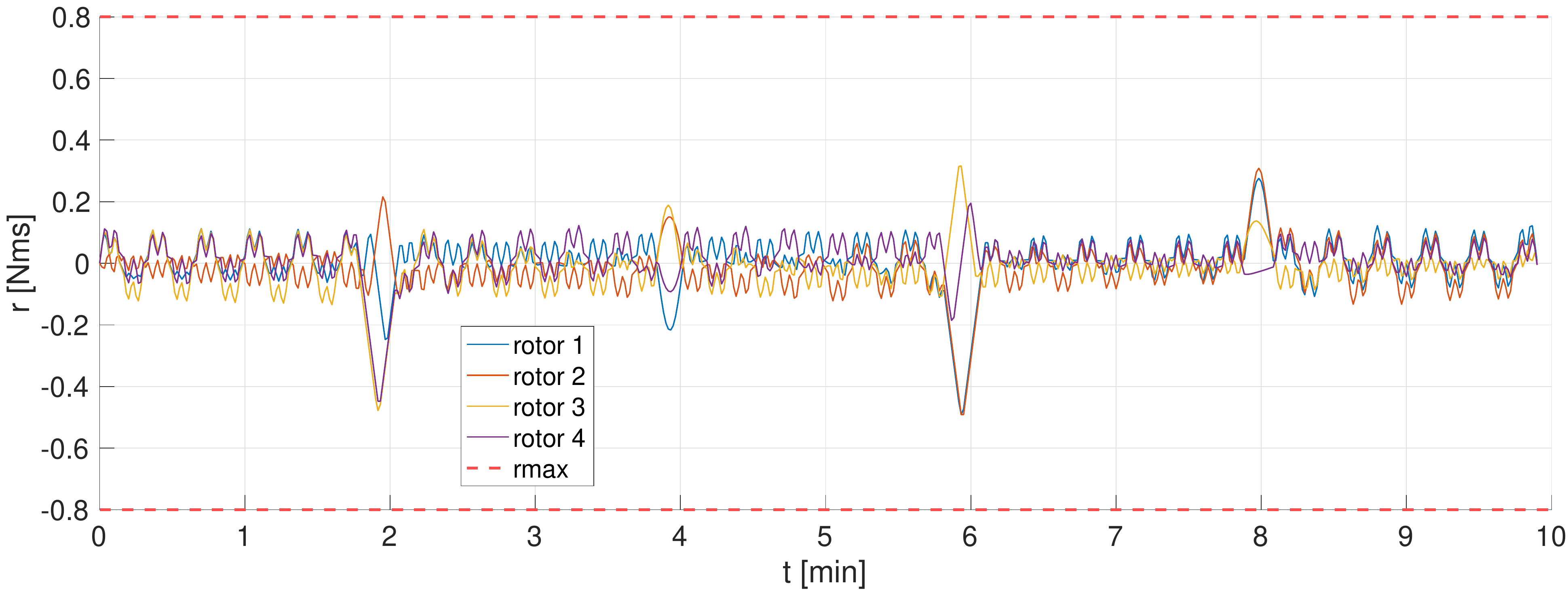}
\caption{Rotor Momentum}
\label{fig:minerror_momentum}
\end{figure}

As shown in Fig. (\ref{fig:minerror_momentum}), the rotor momenta stays within the 0.8 Nms bounds and peaks when conducting large angle maneuvers, such as slewing between regions. Despite the large distances between certain pairs of regions, the rotor momenta do not saturate. This suggests that the reference spacecraft is capable of slewing between regions with much larger ground separations before the rotor momenta saturate. 

\begin{figure}[htbp!]
\centering
\includegraphics[width=0.70\textwidth]{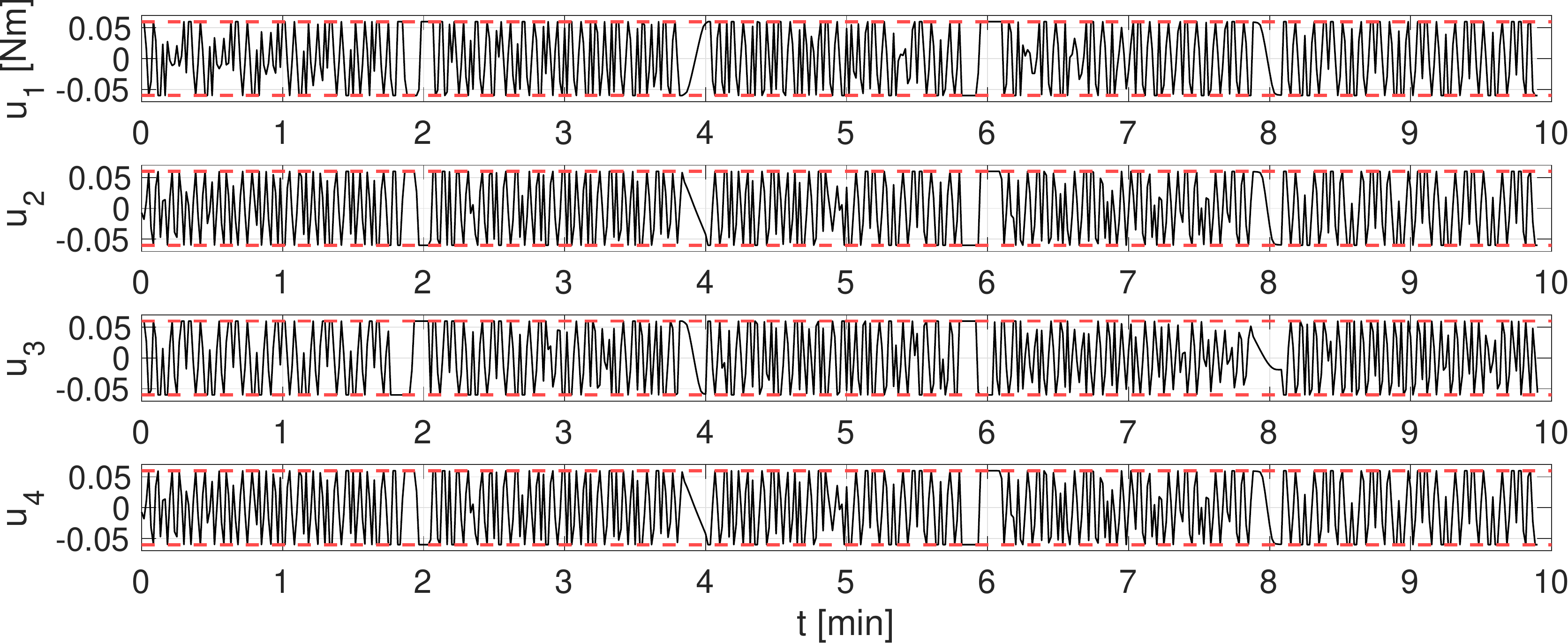}
\caption{Rotor Torque}
\label{fig:minerror_torque}
\end{figure}

Despite the control effort penalty in the objective (\ref{eqn:OPTB_objective}), which encourages minimal control effort, we observe that the rotor torque trajectories saturate at the 0.06 Nm bounds in Fig. (\ref{fig:minerror_torque}) in a ``bang-bang,'' minimum-time fashion. This suggests that the spacecraft is working at its performance limits to achieve an aggressive pointing schedule. We expect actuator saturation to occur for schedules that maximize the number of observed grid points in a certain amount of time.

The rotor momentum and torque trajectories inform us of the reference spacecraft's agile performance limits. Schedules requiring large-angle slew maneuvers may saturate rotor momentum bounds, limiting how far the spacecraft’s pointing vector can span in a given amount of time. Schedules requiring rapid slewing between attitude points may saturate rotor torque bounds. Roughly speaking, for a given spacecraft design (i.e., mass moment-of-inertia, actuator configuration), rotor momentum and torque bounds dictate spatial range-of-motion and speed, respectively.

\begin{figure}[htbp!]
\centering
\includegraphics[width=0.70\textwidth]{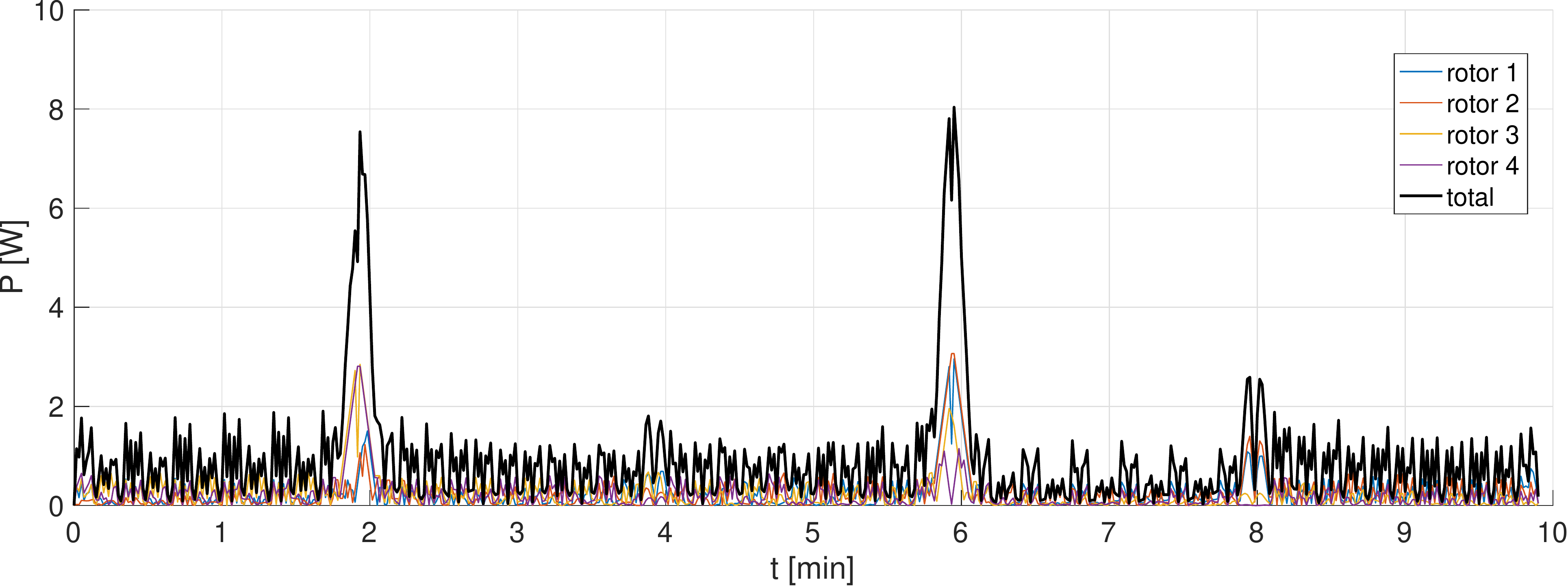}
\caption{ADCS Instantaneous Power drawn}
\label{fig:minerror_power}
\end{figure}

Similar to the body angular velocity and rotor momentum trajectories, we observe in Fig. (\ref{fig:minerror_power}) that the total instantaneous power drawn by the rotors (i.e., ADCS) peaks when slewing between regions. We recall that if maximum power bounds must be enforced on the ADCS, we may add constraint (\ref{eqn:powerconstraint}) in our problem formulation. 

\begin{figure}[htbp!]
\centering
\includegraphics[width=0.70\textwidth]{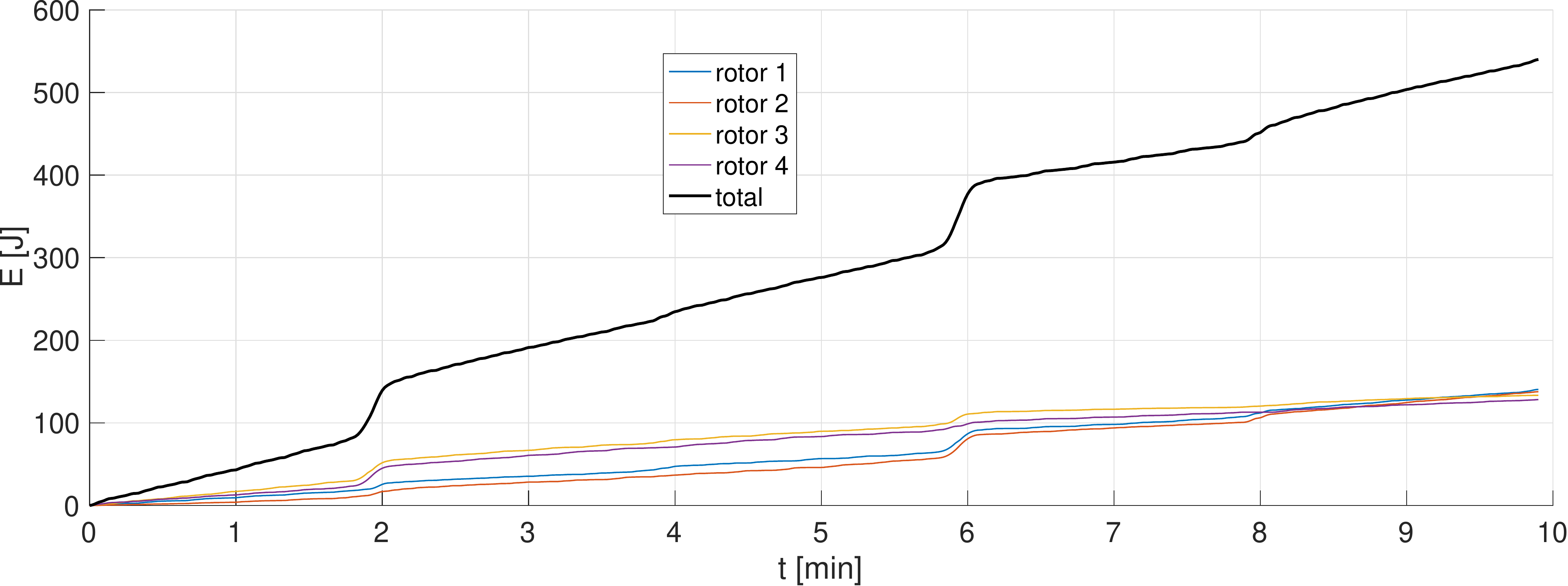}
\caption{ADCS Cumulative Energy consumed}
\label{fig:minerror_energy}
\end{figure}

In Fig. \ref{fig:minerror_energy}, we note that the ADCS requires approximately 550 J to conduct this ten-minute, minimum-effort multi-target pointing trajectory. If this trajectory is too energetically expensive, we may increase the relative control effort weight $\rho$ in the problem objective (\ref{eqn:OCPB_objective}) to further penalize energy consumption. However, the choice may come at the expense of increased attitude error. If a specific energy bound must be satisfied, we may explicitly enforce it by adding constraint  (\ref{eqn:energyconstraint}). In the case that a feasible trajectory with an energy upper bound cannot be found, the attitude pointing schedule may need to be relaxed by observing few regions and/or GPs.

\subsection{Open-loop versus Closed-loop Performance}  \label{AppC}

Using the solution to the \textit{Minimum-Effort Multi-Target Pointing OPT}, we can apply the reference torque trajectory $\{ \mathbf{u}_k^{\scriptscriptstyle OL} := \bar{\mathbf{u}}_k \}_{k=1}^{K}$ as a linearly-interpolated, open-loop control command in simulation with the dynamics described in (\ref{eqn:nonlinsystem}) and parameters listed in Table (\ref{tab:params}). We observe that the resulting state trajectory matches the solution $\{ \bar{\mathbf{q}}_k, \ \ \bar{\boldsymbol{\omega}}_k, \ \ \bar{\mathbf{r}}_k \}_{k=1}^{K}$ with negligible error, signifying that the \textit{OPT} approximation, solved using sequential convex programming, has produced a feasible solution for the original \textit{OCP} formulation. However, in the presence of external disturbances, parameter uncertainty or unmodeled dynamics, a feedback control law must be applied to track the desired attitude trajectory. 

We design an optimal tracking controller based on the solution to the LQR problem (see Appendix) and use the resulting control law as feedback term to regulate deviations from the desired trajectory. The rotor torque command that we apply to (\ref{eqn:nonlinsystem}) consists of both the feedforward and feedback terms: $\{ \mathbf{u}_k^{\scriptscriptstyle CL} := \bar{\mathbf{u}}_k + \delta \mathbf{u}_k \}_{k=1}^{K-1}$, where $\delta \mathbf{u}$ is a time-varying state feedback law.
 
To compare open-loop versus closed-loop performance in presence of model mismatch, we perturb the inertia matrix used in simulation. While we assume a nominal inertia matrix $J$ in the OCP formulation and the LQR tracking control law design, we use a significantly perturbed $\tilde{J}$ in simulation:
\begin{align}
	J = \begin{bmatrix} 8.5 & 0.0 & 0.0 \\ 0.0 & 8.5 & 0.0 \\ 0.0 & 0.0 & 6.0 \end{bmatrix} \ \text{kg$\cdot$m$^2$} \hspace{1.0cm} \tilde{J} = \begin{bmatrix} 15.0 & \text{-}1.0 & \phantom{\text{-}}2.0 \\ \text{-}1.0 & \phantom{\text{-}}7.0 & \text{-}3.0 \\ \phantom{\text{-}}2.0 & \text{-}3.0 & \phantom{\text{-}}9.0 \end{bmatrix} \ \text{kg$\cdot$m$^2$}
\end{align}

\begin{figure}[htbp!]
\centering
\includegraphics[width=0.70\textwidth]{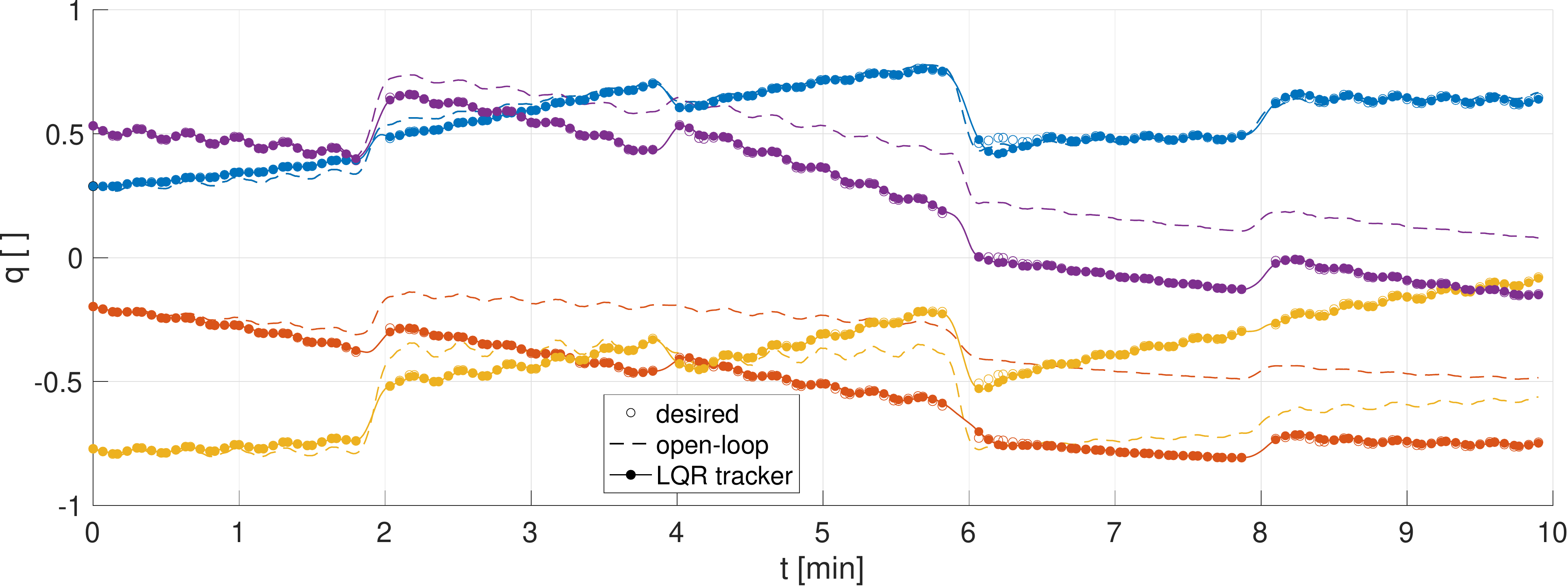} \\
\caption{LQR closed-loop and open-loop quaternion trajectories}
\label{fig:minerror_quaternion_OLvsLQR}
\end{figure}

\begin{figure}[htbp!]
\centering
\includegraphics[width=0.70\textwidth]{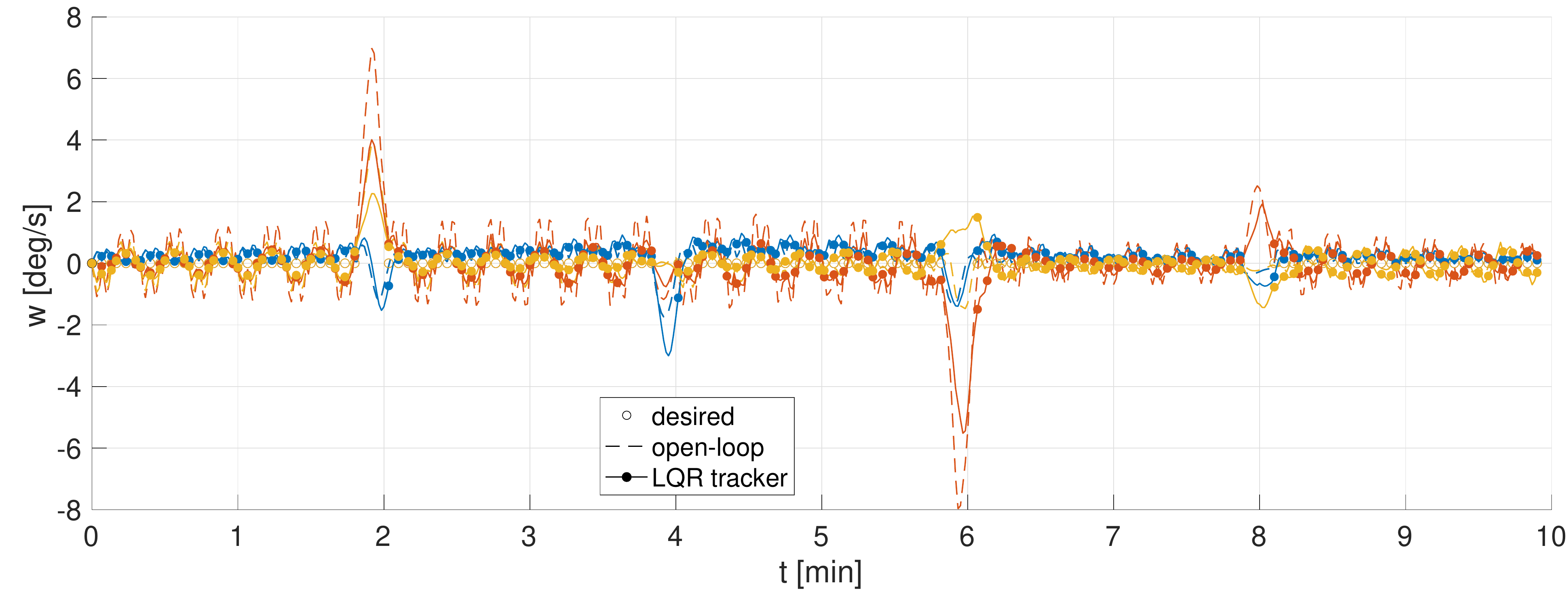} \\
\caption{LQR closed-loop and open-loop angular velocity trajectories}
\label{fig:minerror_omega_OLvsLQR}
\end{figure}

As illustrated in Fig. \ref{fig:minerror_quaternion_OLvsLQR}, applying the optimal torque reference trajectory in open-loop produces the dashed quaternion trajectory, which fails to pass through the desired quaternion points specified by our schedule. However, when using the LQR tracking controller, the closed-loop trajectory is able to pass through most of the desired quaternion points. In Fig. \ref{fig:minerror_omega_OLvsLQR} we confirm that the amplitude of the angular velocity peaks during inter-region slewing are reduced. 

\begin{figure}[htbp!]
\centering
\includegraphics[width=0.85\textwidth]{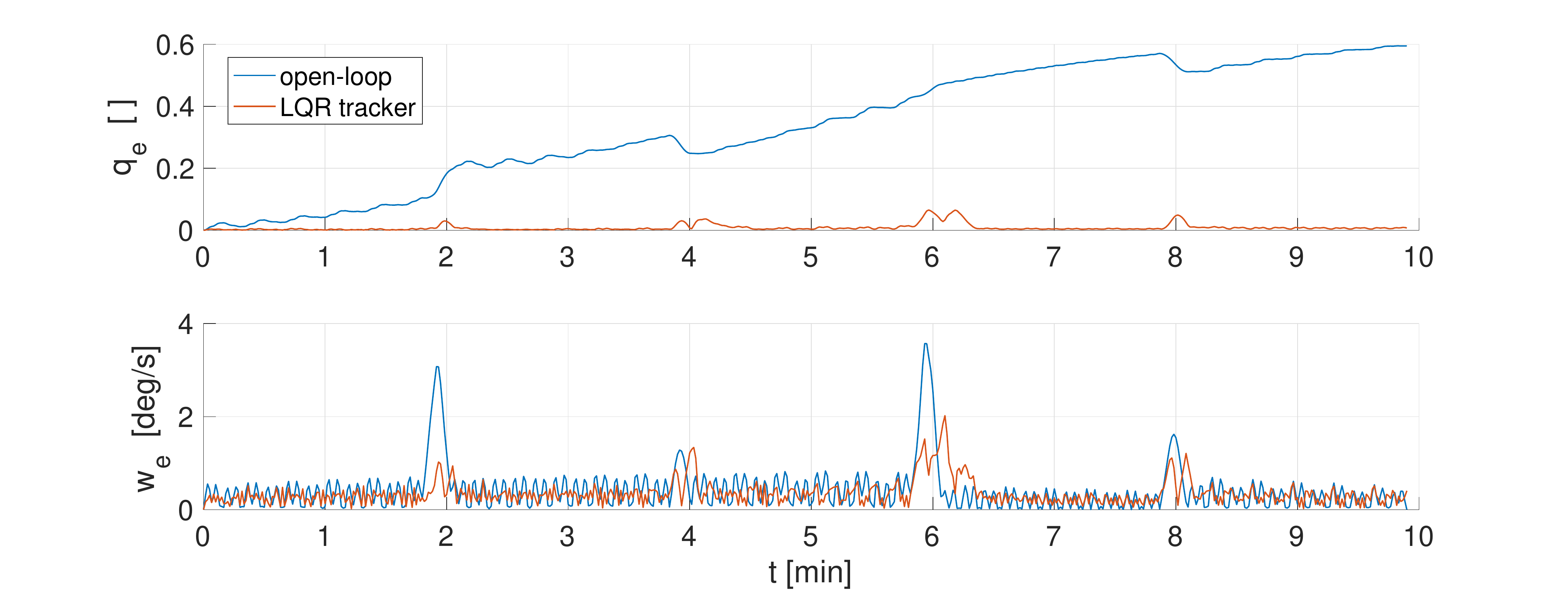} \\
\caption{(Top) Quaternion error, (Bottom) Angular velocity error}
\label{fig:minerror_errors}
\end{figure}

In Section \ref{AppB}, we use the performance metrics (\ref{eqn:qmaxmetrics})-(\ref{eqn:wavgmetrics}) to measure the error between the optimized reference trajectory and the desired attitude schedule. In this section, we use the same metrics to compute the error of the closed-loop (or open-loop) simulated trajectory with respect to the reference trajectory. 
\begin{table}[!htbp]
\centering
\begin{tabular}{c | c c | c c}
\toprule
\multicolumn{1}{c}{} & \multicolumn{2}{c}{Quaternion Error}  & \multicolumn{2}{c}{Angular Velocity Error}  \\
\midrule
Control Strategy    & $q_{e}^{max}$ & $q_{e}^{avg}$ & $\omega_{e}^{max}$ & $\omega_{e}^{avg}$ \\
\hline
Open-loop    & 0.5947 & 0.3417 & 3.5724 & 0.4116   \\
LQR tracker & 0.0653 & 0.0085 & 2.0227 & 0.3515  \\
\bottomrule
\end{tabular}
\caption{Simulation trajectory error with respect to reference trajectory}
\label{tab:trajopt_vs_desired_errors}
\end{table}
As shown in Table \ref{tab:trajopt_vs_desired_errors}, the LQR-based, closed-loop trajectories result in smaller error across all four performance metrics. In particular, the maximum point-wise quaternion (i.e., attitude) error is reduced by an order of magnitude when using the tracking controller. In Fig. \ref{fig:minerror_errors}, we plot the quaternion and angular velocity errors as functions of the simulation time. We observe that the maximum point-wise error in the open-loop quaternion trajectory coincides with the final point in the trajectory. Since the error accumulates over time, we can infer that schedules prescribed for long durations of time will result in large attitude errors if optimal torque commands are applied in open-loop. In contrast, the LQR tracker significantly reduces error in the closed-loop quaternion trajectory. Furthermore, both the maximum and average angular velocity error is reduced with the LQR tracker. Further reduction in the angular velocity error may be achieved (at the expense of either increased quaternion error or control effort) by tuning the LQR weight matrices (\ref{eqn:LQRobjective}).

\section{Conclusion}
We have formulated two optimal control problems, the Minimum-Time Slew OCP and the Minimum-Effort Multi-Target Pointing OCP, that can be applied to high-fidelity spacecraft models with practical mission objectives and constraints. The 3-DOF, rotational motion of a spacecraft may be modeled with nonlinear gyrostat and actuator dynamics, quaternion kinematics, arbitrary physical parameters, and arbitrary actuator configurations. Constraints such as bounds on rotor torque, rotor momentum, body slew rate, ADCS power and energy are included. Using state-of-the-art techniques in sequential convex programming, we transcribe the non-convex, continuous-time optimal control problems into convex, finite-dimensional optimization problems (OCP$\rightarrow$OPT) to be solved efficiently.

The Minimum-Time Slew OPT accurately estimates the required time and energy to conduct time-optimal, rest-to-rest maneuvers between any two arbitrary orientations. Compared with simulation-based approaches that implement near-optimal, eigenaxis-slew control laws about body-fixed control axes, our trajectory optimization approach is not restricted to eigenaxis slews and can be applied to more general spacecraft models and constraints. The time and energy estimates can be computed on a grid of the space of 3D rotations to produce models for use by a constellation scheduler.

Given a desired schedule of discrete pointing orientations, the Minimum-Effort Multi-Target Pointing OPT produces a dynamically-feasible, continuous reference trajectory that can achieve the pointing schedule with minimal effort. To mitigate the effects of model mismatch or external disturbances, a trajectory-tracking controller is used to regulate deviations from the trajectory. Compared to existing approaches that step from one discrete reference set-point to another, our method allows us to design the entire multi-target pointing trajectory as a single unabridged maneuver.
We have applied both problems formulations to an example multi-target observation mission and reference spacecraft, demonstrating their potential use by a remote sensing constellation scheduler.

The two problem formulations are complementary and may significantly elevate the performance of agile, remote sensing satellite constellations.
With more accurate estimates of the minimum time and energy required to conduct particular reorientations, a constellation scheduler is better informed to design aggressive multi-target pointing schedules that maximize constellation utility (e.g., quantity or science value of observations). The satellites of the constellations may then execute the schedules by planning feasible attitude trajectories that explicitly consider state and input constraints. Furthermore, given recent advances in the computing power of small satellite on-board computers, real-time solver implementations allow a constellation to respond quickly to changes in the environment or mission objectives. 

Future work will include a rigorous comparison of our proposed method to existing approaches in the presence of model uncertainty, disturbances, and the use of state estimation. We will also study other applications beyond ground target observation, including the tracking of cloudbows and aerosols in the atmosphere. Finally, our attitude trajectory planning approach will be integrated with an optimization-based scheduler so we may compare its performance to the manually-designed, baseline schedule used in this paper.

\section*{Appendix : Linear Quadratic Trajectory Tracking Controller Design} \label{Appendix}

We use the closed-form solution of the (unconstrained) finite-horizon, discrete-time LQR problem \cite{Athans} - \cite{Anderson} to design a time-varying, state feedback law that regulates deviation from desired trajectory in closed-loop simulation. Due to the intrinsic unit-norm constraint on the quaternion, we convert the error-quaternion to the Euler axis and Angle representation with a transformation $\boldsymbol{\phi} = h(\bar{\mathbf{q}}^+ \mathbf{q} )$ in our tracking controller implementation. Details on the transformation may be found in \cite{Hughes}-\cite{Markley}. 

Using error variables $\boldsymbol{\phi}$, $\delta \boldsymbol{\omega} := (\boldsymbol{\omega} - \bar{\boldsymbol{\omega}})$ and $\delta \mathbf{r} := (\mathbf{r} - \bar{\mathbf{r}})$ to represent deviations from the desired trajectory:
\begin{equation}
  \delta {\mathbf{x}}(t) := [ \boldsymbol{\phi}(t)^\top \ \ \delta \boldsymbol{\omega}(t)^\top \ \ \delta \mathbf{r}(t)^\top ]^\top \quad , \hspace{1.0cm} \delta \mathbf{u}(t) := \mathbf{u}(t) - \bar{\mathbf{u}}(t)  \ ,
\end{equation}
the error system dynamics are:
\begin{equation} \label{eqn:errordyn}
	\dot{\delta {\mathbf{x}(t)}} = \hat{A}(t) \delta \mathbf{x}(t) + \hat{B}(t) \delta \mathbf{u}(t) \ ,
\end{equation}
\begin{equation} \label{eqn:errordynAB}
	\hat{A}(t) :=  \begin{bmatrix} -[\bar{\boldsymbol{\omega}}(t)]^\times & \text{I} & \mathbf{0} \\ \mathbf{0} & -J^{-1}\left( [\bar{\boldsymbol{\omega}}(t)]^\times J - \left[ J \bar{\boldsymbol{\omega}}(t) + A_r \bar{\mathbf{r}}(t)  \right]^\times \right)  & -J^{-1} [\bar{\boldsymbol{\omega}}(t)]^\times A_r \\ \mathbf{0} & \mathbf{0} & \mathbf{0}\end{bmatrix}  \ , \quad \hat{B}(t) := \begin{bmatrix}  \mathbf{0} \\ -J^{-1} A_r \\ \text{I} \end{bmatrix} \ ,
\end{equation}
where $\mathbf{0}$ and I are the $3 \times 3$ null and identity matrices, respectively, and we use the skew-symmetric matrix operator $[ \ ]^\times$ described in (\ref{eqn:skewmats}). 
In the model above, note that we have approximated the error-quaternion kinematics using the Euler axis and angle representation. Assuming a sample time of $\Delta t = \frac{t_f}{K-1}$, we discretize the error dynamics with a zero-order hold on $\delta \mathbf{u}$. The attitude trajectory tracking problem is formulated as:
  \begin{align}
  &\minimize_{ \left\{ \delta {\mathbf{x}}_k, \delta\mathbf{u}_k \right\}_{\scriptscriptstyle k=1}^{\scriptscriptstyle K}} \hspace{0.5cm}  \delta {\mathbf{x}}_K^\top Q_K \delta {\mathbf{x}}_K + \sum_{k=1}^{K-1} \ \ \left\{  \delta {\mathbf{x}}_k^\top Q_k \delta {\mathbf{x}}_k +  \delta\mathbf{u}_k^\top R \delta\mathbf{u}_k \right\} \label{eqn:LQRobjective} \\
  &\ \ \ \text{s.t.} \hspace{1.5cm} \delta {\mathbf{x}}_{k+1} = \hat{A}_{k}\delta {\mathbf{x}}_{k} + \hat{B}_{k}\delta\mathbf{u}_k  \hspace{0.5cm} \forall \ k = 1, \ldots, K- 1
\end{align}
with weight matrices designed so that state trajectory deviations at the observation points can be penalized more heavily using a scaling term $\alpha$:
\begin{equation} \label{eqn:Qmat}
  Q_k :=  
    \begin{cases}
    	\phantom{\alpha \cdot} \ \ Q ,		    & k \not\in \mathbb{K} \\
    	\alpha \cdot Q ,              &  k \in \mathbb{K}
    \end{cases}       
    \hspace{1.0cm} \text{where} \ \ Q \succcurlyeq 0 \ , \ R \succ 0 \ , \alpha > 1 \ .
\end{equation}
The problem formulation admits a unique closed-form solution with feedback control law $\{ \delta\mathbf{u}_k := - \mathbf{K}_k \delta {\mathbf{x}}_k \}_{k=1}^{K-1}$, where:
\begin{align}
	\mathbf{K}_k &= (R_k + \hat{B}_k^\top P_{k+1} B_k)^{\text{-}1} \hat{B}_k^\top P_{k+1} \hat{A}_k \\
	\intertext{and $P_k$ is found recursively from $P_K = Q_K$ using the following discrete-time dynamic Riccati equation:} 
	P_k &= \mathbf{K}_k^\top R_k \mathbf{K}_k + (\hat{A}_k - \hat{B}_k \mathbf{K}_k)^\top P_{k+1} (\hat{A}_k - \hat{B}_k \mathbf{K}_k) + Q_k \ .
\end{align}

\section*{Acknowledgments}
This project has been funded and supported by NASA’s Earth Science Technology Office via the Advanced Information Systems Technology grant. We thank the Orbital R\&D Team at Planet Labs for providing representative parameter values for the reference satellite used in this paper and for sharing insight on the state-of-the-practice in satellite attitude control.

\end{document}